\documentclass[prb,floatfix,twocolumn,showpacs,amsmath,amssymb]{revtex4}
\usepackage{graphics,lscape,graphicx,epsfig,amsmath}
\allowdisplaybreaks
\newcommand{\wt}[1]{\ensuremath{\widetilde{#1}}}
\newcommand{\Pj}[2]{\ensuremath{|#1\rangle\langle #2 |}}

\newcommand{\be}{\begin{equation}}
\newcommand{\ee}{\end{equation}}
\newcommand{\bea}{\begin{eqnarray}}
\newcommand{\eea}{\end{eqnarray}}
\newcommand{\ba}{\begin{eqnarray*}}
\newcommand{\ea}{\end{eqnarray*}}
\newcommand{\dagga}{{\phantom{\dagger}}}
\newcommand{\bR}{\mathbf{R}}

\newcommand{\bk}{\mathbf{k}}

\newcommand{\bRp}{\mathbf{R'}}

\newcommand{\Tr}{\mathrm{Tr}}
\begin{document}

\title{Gutzwiller description of non-magnetic Mott insulators: a dimer lattice model.}

\author{Michele Fabrizio,$^{1,2}$}
\affiliation{
$^1$ International School for Advanced Studies (SISSA) and CNR-INFM-Democritos National Simulation Centre, Via Beirut 2-4, 
I-34014 Trieste, Italy \\
$^2$ The Abdus Salam International Centre for Theoretical Physics (ICTP), P.O. Box 586, I-34014 Trieste, Italy}

\date{\today}

\begin{abstract}
We introduce a novel extension of the Gutzwiller variational wavefunction able to deal with 
insulators that escape any mean-field like description, as for instance non-magnetic insulators. 
As an application, we study the Mott transition from a paramagnetic metal into a non-magnetic Peierls, or valence-bond, 
Mott insulator. We analyze this model by means of our Gutzwiller wavefunction analytically in the limit of large coordination lattices, 
where we find that: (1) the Mott transition is first order; (2) the Peierls gap is large in the Mott insulator, although it is mainly 
contributed by the electron repulsion; (3) singlet-superconductivity arises around the transition.  
\end{abstract}

\pacs{71.10.-w, 71.10.Fd, 71.30.+h}

\maketitle

\section{Introduction}
Among the theoretical tools devised to deal with strongly correlated metals close to a Mott metal-to-insulator 
transition (MIT), the simplest one likely is the variational approach introduced 
in the 60ths' by Gutzwiller~\cite{Gutzwiller1,Gutzwiller2} to describe itinerant ferromagnetism and narrow band conductors.  
In its original version, the Gutzwiller variational wavefunction has the form 
\be
|\Psi_G\rangle = \mathcal{P}\, |\phi\rangle = \prod_\bR\, \mathcal{P}_\bR\, |\phi\rangle,
\label{GWF-old}
\ee
where $|\phi\rangle$ is an uncorrelated wavefunction for which 
Wick's theorem holds, 
$\mathcal{P}_\bR$ an operator at site $\bR$, and both $|\phi\rangle$ and $\mathcal{P}_\bR$ have to be 
determined by minimizing the variational energy. The role of the operator $\mathcal{P}_\bR$ is to modify, 
according to the on-site interaction, 
the weights of the local electronic configurations with respect to their values in the uncorrelated wavefunction.   

In spite of its simplicity, the Gutzwiller wavefunction is quite effective in capturing physical 
properties that supposedly identify 
strongly correlated metals, as for instance the large increase of the effective mass.~\cite{brinkman&rice} 
However, since the dependence upon the distance $|\bR - \bRp|$ of inter-site correlations are still determined by the 
uncorrelated wavefunction, while the local operators $\mathcal{P}_\bR$ just affect the amplitudes, 
the Gutzwiller wavefunction can describe a Mott 
insulator either if $\mathcal{P}_\bR$ suppresses completely charge 
fluctuations, that provides a very poor description of an insulator, or 
if $|\phi\rangle$ itself is insulating. The latter case can be stabilized within the original Gutzwiller approach only 
when $|\phi\rangle$ is an admissible Hartree-Fock solution of the Hamiltonian. As 
an example let us consider a single band model at half-filling, for instance the Hubbard model
\[
\mathcal{H} = -\sum_{\bR\bRp,\sigma}\,t_{\bR\bRp}\,c^\dagger_{\bR,\sigma}c^\dagga_{\bRp,\sigma} 
+ U\sum_\bR\, n_{\bR,\uparrow}\, n_{\bR,\downarrow},
\]
where $c^\dagger_{\bR,\sigma}$ and $c^\dagga_{\bR,\sigma}$ creates and annihilates, respectively, an electron 
with spin $\sigma=\uparrow,\downarrow$ at site $\bR$ and $n_{\bR,\sigma} = c^\dagger_{\bR,\sigma}c^\dagga_{\bR,\sigma}$. 
This Hamiltonian admits at the mean-field level two possible phases, one paramagnetic,  
$\langle n_{\bR,\uparrow}\rangle = \langle n_{\bR,\downarrow}\rangle$, and the other magnetic, 
$\langle n_{\bR,\uparrow}\rangle \not = \langle n_{\bR,\downarrow}\rangle$. The latter is the only one that can 
eventually describe an insulator.   
In the Hubbard model the action of the operator $\mathcal{P}$ is to increase the weight of singly occupied sites at expenses of 
doubly occupied and empty sites, in order to minimize the Coulomb repulsion $U$. 
Evidently, even when the repulsion is very strong, hence the model is a Mott insulator, a realistic wavefunction 
should still allow for charge fluctuations responsible for the super-exchange, that survives even deep inside the Mott phase. 
However, any paramagnetic uncorrelated wavefunction, for instance the Fermi sea, is unable to generate any super-exchange 
and necessarily leads to a non-realistic Mott insulator where configurations with empty or doubly occupied sites 
are fully suppressed.~\cite{brinkman&rice} The only way to generate super-exchange is to 
assume a  magnetically ordered $|\phi\rangle$, which is also the only insulating wavefunction 
accessible within Hartree-Fock. 
However a magnetic state might not always be the right choice, especially if magnetism is sufficiently frustrated. 

Recently, an improved version of the Gutzwiller wavefunction has been 
proposed,~\cite{Capello1} in which additional 
inter-site correlations are provided by density-density Jastrow factors, namely 
\be
|\Psi_G\rangle \rightarrow \exp{\Big[-\sum_{\bR \bRp}\, v_{\bR,\bRp}\, n_\bR\,n_\bRp\Big]}\; |\Psi_G\rangle ,
\label{GWF-Manuela}
\ee
where $n_\bR$ is the site $\bR$ occupation number and $v_{\bR,\bRp}$ variational parameters. 
This novel class of wavefunctions has the capability to disentangle charge from other degrees of freedom, 
hence is more suitable to capture Mott localization, as it has indeed been shown.~\cite{Capello1,Capello2,Capello4}
However, unlike the conventional Gutzwiller 
wavefunction (\ref{GWF-old}), the Gutzwiller-Jastrow wavefunction (\ref{GWF-Manuela}) can only be dealt with 
numerically by variational Monte Carlo, which is inherently limited to finite-size systems, 
albeit quite large.~\cite{Sorella-VMC}

An alternative approach, that is closely related to recently proposed extensions of Dynamical Mean Field Theory 
(DMFT) from the original single-site formulation~\cite{DMFT} to 
a cluster one~\cite{CDMFT-Lichtenstein,CDMFT-Senechal,CDMFT-Potthoff,CDMFT-Jarrell,CDMFT-Kotliar},  
is to consider a variational 
wavefunction of the same form as (\ref{GWF-old}) but defined on a lattice with non-primitive unit cells. 
In this case, the operator $\mathcal{P}_\bR$ acts on all the available electronic configurations of the lattice sites belonging to 
the non-primitive cell. 
The advantage is that in this way one may include additional short-range correlations without losing the property of the 
wavefunction to be analytically manageable, at least in infinite-coordination lattices.  
The obvious disadvantage is that this wavefunction could bias the variational solution towards translational-symmetry breaking. 

Within this scheme, the variational problem becomes generically equivalent to optimize a Gutzwiller wavefunction 
for a multi-band Hamiltonian. 
There have been recently an amount of attempts to extend the Gutzwiller wavefunction 
to multi-orbital models that include further complications like for instance Coulomb 
exchange~\cite{Gebhard,Attaccalite,wang:2006,Ferrero}.  
In this paper we introduce a further extension that is capable to generate 
inter-site correlations as the super-exchange for paramagnetic wavefunctions, otherwise missed by the conventional 
Gutzwiller approach. This novel class of wavefunctions also allows to explore new kinds of variational solutions. 
Specifically, there are interesting examples of correlated models where 
the Mott insulating phase escapes any Hartree-Fock mean-field treatment, in other words can not be represented by a single 
Slater determinant. A very simple case, that we will explicitly consider throughout this work,  
is a Peierls insulator, namely a short-range valence-bond crystal, in which pairs of nearest neighbor sites 
are strongly bound into a singlet configuration, leading to a state that is simply a collection of spin-singlets.  
Such a Mott insulating state is not accessible by Hartree-Fock theory, just because each singlet is itself not 
expressible as a Slater determinant, nor by the conventional Gutzwiller approach, which, as mentioned, 
gives a poor description of paramagnetic insulators.

The paper is organized as follows. In Section~\ref{Sec.I} we present the variational wavefunction and discuss under which 
conditions it can be deal with analytically. In Section~\ref{Sec.II} we discuss how to build up the wavefunction 
in the case in which the basic unit of the lattice model is a dimer. Next, in Section~\ref{Sec.III}, we solve the 
variational problem for a specific lattice model of dimers. Conclusions are given in Section~\ref{Sec.IV}.

\section{The variational wavefunction}
\label{Sec.I}

In this Section, we introduce an extension of the Gutzwiller wavefunction (\ref{GWF-old}) which 
is particularly convenient to perform analytical calculation in the limit of infinite-coordination lattices.\cite{Gebhard}
Let us consider a generic multi-band Hamiltonian. Each lattice site $\bR$ contains several orbitals that give rise to 
a bunch of electronic configurations which we denote individually as $|\Gamma;\bR\rangle$. The most general operator 
$\mathcal{P}_\bR$ can be chosen of the form:
\be
\mathcal{P}_\bR = \sum_{\Gamma \Gamma'}\, \lambda(\bR)_{\Gamma\Gamma'}\, \Pj{\Gamma;\bR}{\Gamma';\bR},
\ee
where $\lambda(\bR)_{\Gamma\Gamma'}$ are variational parameters. In general $\mathcal{P}_\bR$ needs not to be hermitean, 
namely for $\Gamma\not = \Gamma'$ it is not required that $\lambda(\bR)_{\Gamma\Gamma'}^* = \lambda(\bR)_{\Gamma'\Gamma}$. 
Indeed, as we shall see, the non-hermitean character plays a very important role. We further assume that the 
Wick's theorem holds for the uncorrelated wavefunction, hence that $|\phi\rangle$ is either a Slater determinant 
or a BCS wavefunction. 

It was realized by B\"unemann, Weber and  Gebhard\cite{Gebhard} that average values of operators on the Gutzwiller 
wavefunction (\ref{GWF-old}) can be analytically computed in infinite coordination 
lattices provided the following two constraints are imposed on $\mathcal{P}_\bR$:
\bea
\langle \phi|\, \mathcal{P}_\bR^\dagger \,\mathcal{P}_\bR^\dagga \, |\phi\rangle &=& 
\langle \phi |\phi\rangle = 1,\label{cond-1}\\
\langle \phi|\, \mathcal{P}_\bR^\dagger \,\mathcal{P}_\bR^\dagga \, \mathcal{C}_\bR\, |\phi\rangle &=& 
\langle \phi|\,  \mathcal{C}_\bR\, |\phi\rangle
,\label{cond-2}
\eea
where $\mathcal{C}_\bR$ is the local single-particle density-matrix operator, with elements 
$c^\dagger_{\bR,\alpha}\,c^\dagga_{\bR,\beta}$ and $c^\dagger_{\bR,\alpha}\,c^\dagger_{\bR,\beta}$, 
$\alpha$ labeling single-particle states, while $c^\dagger_{\bR,\alpha}$ and $c^\dagga_{\bR,\alpha}$
create and annihilate, respectively, an electron at site $\bR$ in state $\alpha$. 

The first constraint, Eq.~(\ref{cond-1}),  
does not actually limit the variational freedom, since $\mathcal{P}_\bR$ is defined up to a normalization factor. 
On the contrary, the latter constraint, Eq.~(\ref{cond-2}), may reduce the variational freedom, 
although it seems not in a relevant manner, at least in all cases that we have so far investigated. 
We notice that Eq.~(\ref{cond-2}) is not the same as imposing 
\be
\langle \phi|\, \mathcal{P}_\bR^\dagger \,\mathcal{C}_\bR\,\mathcal{P}_\bR^\dagga \,  |\phi\rangle = 
\langle \phi|\,  \mathcal{C}_\bR\, |\phi\rangle,
\label{cond-3}
\ee
unless $\mathcal{P}_\bR^\dagga$ commutes with $ \mathcal{C}_\bR$, which is a further constraint to be 
imposed on $\mathcal{P}_\bR$. This actually is the only case that has been hitherto considered, 
see e.g. Refs.~\onlinecite{Gebhard} and \onlinecite{Attaccalite}. However, as we shall see, there are interesting models 
which force to abandon the supplementary condition (\ref{cond-3}), which is anyway unnecessary.~\cite{Ferrero}   

\begin{figure}[t]
\includegraphics[width=8cm]{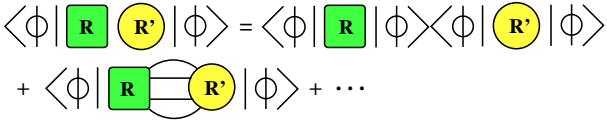}
\caption{Graphical representation of the average on $|\phi\rangle$ 
of $\mathcal{P}_\bR^\dagger\, \mathcal{P}^{~}_\bR$, drawn as 
a box, times a generic multi-particle operator at site $\bRp$, drawn as a circle. Lines that join the two operators 
represent the average of two single-fermion operators, one at $\bR$ and the other at $\bRp$.
The dots include all terms where the two sites are joined by more than four lines. The important thing to notice is the 
absence of terms in which the two sites are connected by two lines.\label{Fig1}}
\end{figure}

By means of Wick's theorem, the left-hand side of (\ref{cond-2}) includes a disconnected term
\[
\langle \phi|\, \mathcal{P}_\bR^\dagger \,\mathcal{P}_\bR^\dagga \, |\phi\rangle\,
\langle \phi|\,  \mathcal{C}_\bR\, |\phi\rangle 
= \langle \phi|\,  \mathcal{C}_\bR\, |\phi\rangle ,
\]
where the right-hand side follows from (\ref{cond-1}), plus connected terms that are obtained by selecting 
in all possible ways a pair of single-fermion operators from $\mathcal{P}_\bR^\dagger \,\mathcal{P}_\bR^\dagga$, 
averaging on $|\phi\rangle$ what remains, and finally averaging the two single-fermion operators with 
those of $\mathcal{C}_\bR$. Therefore, imposing (\ref{cond-2}) means that the sum of all connected terms vanishes, 
whatever is the element of the single-particle density-matrix. In other words, the operator that is left 
after taking out from $\mathcal{P}_\bR^\dagger \,\mathcal{P}_\bR^\dagga$ any pair of single-fermion operators 
has null average on $|\phi\rangle$. In turns, this also implies that, when averaging on $|\phi\rangle$ 
$\mathcal{P}_\bR^\dagger \,\mathcal{P}_\bR^\dagga$ with multi-particle operators at different sites, 
the only connected terms that survive are those that involve four or more single-fermion operators 
of $\mathcal{P}_\bR^\dagger \,\mathcal{P}_\bR^\dagga$, that are represented  
graphically in Fig.~\ref{Fig1} as lines coming out of $\mathcal{P}_\bR^\dagger \,\mathcal{P}_\bR^\dagga$.
This property of $\mathcal{P}_\bR$ turns out to be extremely useful in infinite-coordination lattices. In this limit,   
the contribution to the average value on $|\phi\rangle$ of terms in which more than two fermionic lines come out of 
$\mathcal{P}_\bR^\dagger \,\mathcal{P}_\bR^\dagga$ can be shown to vanish\cite{Gebhard}, which simplifies considerably 
all calculations. For instance, the average value on (\ref{GWF-old}) of any local operator $\mathcal{O}_\bR$ becomes
\be
\langle \phi|\,\mathcal{P}^\dagger\,\mathcal{O}_\bR\,\mathcal{P}\,|\phi\rangle = 
\langle \phi| \,\mathcal{P}_\bR^\dagger\, \mathcal{O}_\bR\,\mathcal{P}_\bR^\dagga |\phi\rangle,
\label{OR}
\ee
which also implies, taking $\mathcal{O}_\bR = 1$, that the variational wavefunction (\ref{GWF-old}) is normalized.
In addition, the average value of the inter-site single-particle density matrix turns out to be
\bea
&& \langle \phi|\,\mathcal{P}^\dagger\,c^\dagger_{\bR,\alpha}\,c^\dagga_{\bRp,\beta}\, \mathcal{P}\,|\phi\rangle
\nonumber \\
&&~~= \langle \phi|\,\mathcal{P}_\bR^\dagger\,c^\dagger_{\bR,\alpha}\,\mathcal{P}_\bR^\dagga\, 
\mathcal{P}_\bRp^\dagger\, c^\dagga_{\bRp,\beta}\, \mathcal{P}_\bRp^\dagga\,|\phi\rangle\nonumber \\
&&~~ = 
\sum_{\gamma\delta}\, \sqrt{Z(\bR)_{\alpha\gamma}^\dagga\,Z(\bRp)_{\beta\delta}^*}
\, \langle \phi|\,c^\dagger_{\bR,\gamma}\,c^\dagga_{\bRp,\delta}\,|\phi\rangle
\nonumber\\
&&~~~ +  
\sum_{\gamma\delta}\, \sqrt{Z(\bR)_{\alpha\gamma}^\dagga\,\Delta(\bRp)_{\beta\delta}^*}
\, \langle \phi|\,c^\dagger_{\bR,\gamma}\,c^\dagger_{\bRp,\delta}\,|\phi\rangle
\nonumber\\
&&~~~ +
\sum_{\gamma\delta}\, \sqrt{\Delta(\bR)_{\alpha\gamma}^\dagga\,Z(\bRp)_{\beta\delta}^*}
\, \langle \phi|\,c^\dagga_{\bR,\gamma}\,c^\dagga_{\bRp,\delta}\,|\phi\rangle
\nonumber\\
&&~~~ +
\sum_{\gamma\delta}\, \sqrt{\Delta(\bR)_{\alpha\gamma}^\dagga\,\Delta(\bRp)_{\beta\delta}^*}
\, \langle \phi|\,c^\dagga_{\bR,\gamma}\,c^\dagger_{\bRp,\delta}\,|\phi\rangle
, \label{cdagger-c}\\
&& \langle \phi|\,\mathcal{P}^\dagger\,c^\dagger_{\bR,\alpha}\,c^\dagger_{\bRp,\beta}\, \mathcal{P}\,|\phi\rangle
\nonumber \\
&&~~= \langle \phi|\,\mathcal{P}_\bR^\dagger\,c^\dagger_{\bR,\alpha}\,\mathcal{P}_\bR^\dagga\, 
\mathcal{P}_\bRp^\dagger\, c^\dagger_{\bRp,\beta}\, \mathcal{P}_\bRp^\dagga\,|\phi\rangle\nonumber\\
&&~~= 
\sum_{\gamma\delta}\, \sqrt{Z(\bR)_{\alpha\gamma}^\dagga\,Z(\bRp)_{\beta\delta}}
\, \langle \phi|\,c^\dagger_{\bR,\gamma}\,c^\dagger_{\bRp,\delta}\,|\phi\rangle
\nonumber\\
&&~~~ +  
\sum_{\gamma\delta}\, \sqrt{Z(\bR)_{\alpha\gamma}^\dagga\,\Delta(\bRp)_{\beta\delta}}
\, \langle \phi|\,c^\dagger_{\bR,\gamma}\,c^\dagga_{\bRp,\delta}\,|\phi\rangle
\nonumber\\
&&~~~ +
\sum_{\gamma\delta}\, \sqrt{\Delta(\bR)_{\alpha\gamma}^\dagga\,Z(\bRp)_{\beta\delta}}
\, \langle \phi|\,c^\dagga_{\bR,\gamma}\,c^\dagger_{\bRp,\delta}\,|\phi\rangle
\nonumber\\
&&~~~ +
\sum_{\gamma\delta}\, \sqrt{\Delta(\bR)_{\alpha\gamma}^\dagga\,\Delta(\bRp)_{\beta\delta}}
\, \langle \phi|\,c^\dagga_{\bR,\gamma}\,c^\dagga_{\bRp,\delta}\,|\phi\rangle
, \label{cdagger-cdagger}
\eea
where the matrices $Z$ and $\Delta$ are determined by inverting the following set of equations
\bea
&&\langle \phi|\,\mathcal{P}_\bR^\dagger\,c^\dagger_{\bR,\alpha}\,\mathcal{P}_\bR^\dagga\,
c^\dagga_{\bR,\beta}\,|\phi\rangle \\
&&~~=  \sum_\gamma\, \sqrt{Z(\bR)_{\alpha\gamma}}\, 
\langle \phi|\,c^\dagger_{\bR,\gamma}\,c^\dagga_{\bR,\beta}\,|\phi\rangle
\nonumber\\
&&~~~ + \sum_\gamma\, \sqrt{\Delta(\bR)_{\alpha\gamma}}\, \langle \phi|\,c^\dagga_{\bR,\gamma}\,c^\dagga_{\bR,\beta}\,|\phi\rangle,
\label{Z-Delta:1}\\
&&\langle \phi|\,\mathcal{P}_\bR^\dagger\,c^\dagger_{\bR,\alpha}\,\mathcal{P}_\bR^\dagga\,
c^\dagger_{\bR,\beta}\,|\phi\rangle \\
&&=~~  \sum_\gamma\, \sqrt{Z(\bR)_{\alpha\gamma}}\, 
\langle \phi|\,c^\dagger_{\bR,\gamma}\,c^\dagger_{\bR,\beta}\,|\phi\rangle
\nonumber\\
&&~~~ + \sum_\gamma\, \sqrt{\Delta(\bR)_{\alpha\gamma}}\, \langle \phi|\,c^\dagga_{\bR,\gamma}\,c^\dagger_{\bR,\beta}\,|\phi\rangle,
\label{Z-Delta:2}.
\eea
Na\"{\i}vely speaking, it is as if, when calculating the inter-site density matrix, a fermionic operator transforms 
effectively into  
\be
c^\dagger_{\bR,\alpha} = \sum_\beta\,\sqrt{Z(\bR)_{\alpha\beta}}\,c^\dagger_{\bR,\beta}
+ \sqrt{\Delta(\bR)_{\alpha\beta}}\,c^\dagga_{\bR,\beta},\label{c-Zc}
\ee
namely that a particle turns into a particle or a hole with probabilities $Z$ and $\Delta$, respectively.
Although all the above expressions are strictly valid only in infinite-coordination lattices, 
it is quite common to use the same formulas also to evaluate average values on the Gutzwiller 
wavefunction in finite-coordination lattices.
This approximation is refereed 
to as the Gutzwiller approximation~\cite{Gutzwiller1,Gutzwiller2,Metzner-Vollhardt-PRL,Metzner-Vollhardt-PRB,Gebhard1}, 
and is known to be equivalent to the saddle point solution within the slave-boson technique.~\cite{Kotliar&Ruckenstein}  

We conclude by noting that the constraint (\ref{cond-2}) turns out to be useful also when the variational 
wavefunction (\ref{GWF-old}) is applied to Anderson impurity models. In this case the operator $\mathcal{P}$ acts only on 
the electronic configurations $|\Gamma\rangle$ of the impurity, namely
\[
\mathcal{P} = \sum_{\Gamma\Gamma'}\,\lambda_{\Gamma\Gamma'}\, \Pj{\Gamma}{\Gamma'}.
\]
If we impose 
\bea
\langle \phi|\, \mathcal{P}^\dagger \,\mathcal{P}^\dagga \, |\phi\rangle &=& 1,\label{cond-1-AIM}\\
\langle \phi|\, \mathcal{P}^\dagger \,\mathcal{P}^\dagga \, \mathcal{C}_{imp}\, |\phi\rangle &=& 
\langle \phi|\,  \mathcal{C}_{imp}\, |\phi\rangle,\label{cond-2-AIM}
\eea
where $\mathcal{C}_{imp}$ is the single-particle density matrix of the impurity, 
then, for any operator of the conduction bath, 
$\mathcal{O}_{bath}$, and because of (\ref{cond-1}) and (\ref{cond-2}), the following result holds
\be
\langle \phi|\, \mathcal{P}^\dagger \,\mathcal{O}_{bath}\, \mathcal{P}^\dagga \, |\phi\rangle
 = \langle \phi|\,\mathcal{O}_{bath}\,|\phi\rangle.
\label{results-AIM}
\ee

\subsection{Some formal definitions}

In order to perform actual calculations, it is convenient to introduce some notations. We define a matrix 
$F_\alpha^\dagga$ with elements 
\[
\left(F_\alpha\right)_{\Gamma_1\Gamma_2} = \langle \Gamma_1;\bR|\, c^\dagga_{\bR,\alpha}\,|\Gamma_2;\bR\rangle,
\]
as well as its hermitean conjugate, $F_\alpha^\dagger$, where we assumed that the definition of the local configurations is the same for all sites. 
It follows that 
\ba
F_\alpha^\dagga\,F_\beta^\dagger + F_\beta^\dagger\,F_\alpha = \delta_{\alpha\beta}\, I,\\
F_\alpha^\dagga\,F_\beta^\dagga + F_\beta^\dagga\,F_\alpha = 0,
\ea
where $I$ is the identity. Next, we introduce the uncorrelated occupation-probability matrix, $P_0(\bR)$, with elements
\be
\left(P_0(\bR)\right)_{\Gamma_1\Gamma_2} = \langle \phi|\Gamma_1;\bR\rangle\, \langle \Gamma_2;\bR|\phi\rangle,\label{formal-def:P0}
\ee
that satisfies
\ba
1 &=& \mathrm{Tr}\left(P_0(\bR)\right),\\
\langle \phi|\,c^\dagger_{\bR,\alpha}\,c^\dagga_{\bR,\beta}\,|\phi\rangle &=& 
\Tr\left(P_0(\bR)\,F_\alpha^\dagger\,F_\beta^\dagga\right),\\
\langle \phi|\,c^\dagger_{\bR,\alpha}\,c^\dagger_{\bR,\beta}\,|\phi\rangle &=& 
\Tr\left(P_0(\bR)\,F_\alpha^\dagger\,F_\beta^\dagger\right).
\ea
Analogously, the variational parameters that define $\mathcal{P}_\bR^\dagga$, $\lambda(\bR)_{\Gamma_1\Gamma_2}$, 
are interpreted as elements of a matrix $\lambda(\bR)$. With these definitions, Eqs.~(\ref{cond-1}) and (\ref{cond-2}) 
become
\ba
\langle \phi|\,\mathcal{P}_\bR^\dagger \, \mathcal{P}_\bR^\dagga\,|\phi\rangle &=&
\Tr\left(P_0(\bR)\,\lambda(\bR)^\dagger\,\lambda(\bR)\right)\\
&=& 1,\\
\langle \phi|\,\mathcal{P}_\bR^\dagger \, \mathcal{P}_\bR^\dagga \,c^\dagger_{\bR,\alpha}\,c^\dagga_{\bR,\beta}\,|\phi\rangle &=& 
\Tr\left(P_0(\bR)\,\lambda(\bR)^\dagger\,\lambda(\bR)\,F_\alpha^\dagger\,F_\beta^\dagga\right)\\
&=& \langle \phi|\,c^\dagger_{\bR,\alpha}\,c^\dagga_{\bR,\beta}\,|\phi\rangle,\\
\langle \phi|\,\mathcal{P}_\bR^\dagger \, \mathcal{P}_\bR^\dagga\,c^\dagger_{\bR,\alpha}\,c^\dagger_{\bR,\beta}\,|\phi\rangle &=& 
\Tr\left(P_0(\bR)\,\lambda(\bR)^\dagger\,\lambda(\bR)\,F_\alpha^\dagger\,F_\beta^\dagger\right)\\
&=& \langle \phi|\,c^\dagger_{\bR,\alpha}\,c^\dagger_{\bR,\beta}\,|\phi\rangle,
\ea 
that suggests to introduce a variational occupation-probability matrix $P(\bR) = P_0(\bR)\,\lambda(\bR)^\dagger\,\lambda(\bR)$ 
with matrix elements 
\be
\left(P(\bR)\right)_{\Gamma_1\Gamma_2} = \sum_{\Gamma_3\Gamma_4}\, 
\left(P_0(\bR)\right)_{\Gamma_1\Gamma_3}\, \lambda(\bR)_{\Gamma_3\Gamma_4}^\dagger\,
\lambda(\bR)_{\Gamma_4\Gamma_2}^\dagga,\label{formal-def:P}
\ee
that must satisfy   
\bea
\mathrm{Tr}\left(P(\bR)\right) &=& 1,\label{formal:cond-1}\\
\Tr\left(P(\bR)\,F_\alpha^\dagger\,F_\beta^\dagga\right)
&=& \langle \phi|\,c^\dagger_{\bR,\alpha}\,c^\dagga_{\bR,\beta}\,|\phi\rangle, \label{formal:cond-2a}\\
\Tr\left(P(\bR)\,F_\alpha^\dagger\,F_\beta^\dagger\right)
&=& \langle \phi|\,c^\dagger_{\bR,\alpha}\,c^\dagger_{\bR,\beta}\,|\phi\rangle, \label{formal:cond-2b},
\eea
Eqs.~(\ref{formal:cond-1}), (\ref{formal:cond-2a}) and (\ref{formal:cond-2b}) replace the constraints 
(\ref{cond-1}) and (\ref{cond-2}). With these definitions, the matrices $Z$ and $\Delta$, see Eqs.~(\ref{Z-Delta:1}) 
and (\ref{Z-Delta:2}), are obtained by solving
\bea
&&\Tr\left(P_0(\bR)\,\lambda(\bR)^\dagger\,F_\alpha^\dagger\,\lambda(\bR)\,F_\beta^\dagga\right) \nonumber\\
&&~=\sum_\gamma\,\sqrt{Z(\bR)_{\alpha\gamma}}\, \Tr\left(P_0(\bR)\,F_\gamma^\dagger\,F_\beta^\dagga\right)\nonumber\\
&&~~ + \sum_\gamma\,\sqrt{\Delta(\bR)_{\alpha\gamma}}\, \Tr\left(P_0(\bR)\,F_\gamma^\dagga\,F_\beta^\dagga\right),
\label{formal-Z-Delta:1}\\
&&\Tr\left(P_0(\bR)\,\lambda(\bR)^\dagger\,F_\alpha^\dagger\,\lambda(\bR)\,F_\beta^\dagger\right) \nonumber\\
&&~=\sum_\gamma\,\sqrt{Z(\bR)_{\alpha\gamma}}\, \Tr\left(P_0(\bR)\,F_\gamma^\dagger\,F_\beta^\dagger\right)\nonumber\\
&&~~ + \sum_\gamma\,\sqrt{\Delta(\bR)_{\alpha\gamma}}\, \Tr\left(P_0(\bR)\,F_\gamma^\dagga\,F_\beta^\dagger\right).
\label{formal-Z-Delta:2}
\eea

The above equations simplify if one uses the natural basis, 
namely the single-particle basis that diagonalizes the density-matrix,
\ba
\langle \phi|\,c^\dagger_{\bR,\alpha}\,c^\dagga_{\bR,\beta}|\phi\rangle &=& n(\bR)_\alpha\,\delta_{\alpha\beta},\\
\langle \phi|\,c^\dagger_{\bR,\alpha}\,c^\dagger_{\bR,\beta}|\phi\rangle &=& 0.
\ea
In this case
\bea
\sqrt{Z(\bR)_{\alpha\beta}} &=& 
\frac{\displaystyle \Tr\left(P_0(\bR)\,\lambda(\bR)^\dagger\,F_\alpha^\dagger\,\lambda(\bR)\,F_\beta^\dagga\right)}
{\displaystyle n(\bR)_\beta},\label{def:Z-natural}\\
\sqrt{\Delta(\bR)_{\alpha\beta}} &=& 
\frac{\displaystyle \Tr\left(P_0(\bR)\,\lambda(\bR)^\dagger\,F_\alpha^\dagger\,\lambda(\bR)\,F_\beta^\dagger\right)}
{\displaystyle 1-n(\bR)_\beta}.\label{def:Delta-natural}
\eea
Moreover, if one constructs the states $|\Gamma;\bR\rangle$ so that $P_0(\bR)$ is diagonal
\[
\left(P_0(\bR)\right)_{\Gamma\Gamma'} = \delta_{\Gamma\Gamma'}\, P_0(\bR;\Gamma),
\]
then 
\be
\left(P(\bR)\right)_{\Gamma_1\Gamma_2} =  
P_0(\bR;\Gamma_1)\,\sum_{\Gamma_3}\, \lambda(\bR)_{\Gamma_1\Gamma_3}^\dagger\,
\lambda(\bR)_{\Gamma_3\Gamma_2}^\dagga,\label{formal-def:P-natural}
\ee

\subsection{The variational problem}

We are now in position to settle up the variational problem. We consider a generic tight-binding 
Hamiltonian 
\bea
\mathcal{H} &=& -\sum_{\bR\bRp}\,\sum_{\alpha\beta}\, 
t^{\alpha\beta}_{\bR\bRp}\, c^\dagger_{\bR,\alpha}c^\dagga_{\bRp,\beta} \nonumber \\
&& + \sum_\bR\,\sum_{\Gamma\Gamma'}\, E(\bR)_{\Gamma\Gamma'}\,\Pj{\Gamma;\bR}{\Gamma';\bR},
\label{generic-Ham}
\eea
where $\alpha$ and $\beta$ stem for spin, orbital and lattice site in the chosen unit cell, and 
the hermitean matrix $E(\bR)$ with elements $E(\bR)_{\Gamma\Gamma'}$ may be also unit-cell dependent. 
The average value of this Hamiltonian on the Gutzwiller wavefunction (\ref{GWF-old}) in the limit of 
infinite coordination lattices or, in finite coordination ones, within the Gutzwiller approximation, is 
\bea
E_{var} &=& -\sum_{\bR\bRp}\,\sum_{\alpha\beta\gamma\delta}\,t^{\alpha\beta}_{\bR\bRp}\, 
\Bigg[\nonumber \\
&&~~~~ \sqrt{Z(\bR)_{\alpha\gamma}^\dagga\,Z(\bRp)_{\beta\delta}^*}
\, \langle \phi|\,c^\dagger_{\bR,\gamma}\,c^\dagga_{\bRp,\delta}\,|\phi\rangle
\nonumber\\
&& ~~~~+  
\sqrt{Z(\bR)_{\alpha\gamma}^\dagga\,\Delta(\bRp)_{\beta\delta}^*}
\, \langle \phi|\,c^\dagger_{\bR,\gamma}\,c^\dagger_{\bRp,\delta}\,|\phi\rangle
\nonumber\\
&& ~~~~+
\sqrt{\Delta(\bR)_{\alpha\gamma}^\dagga\,Z(\bRp)_{\beta\delta}^*}
\, \langle \phi|\,c^\dagga_{\bR,\gamma}\,c^\dagga_{\bRp,\delta}\,|\phi\rangle
\nonumber\\
&& ~~~~+
\sqrt{\Delta(\bR)_{\alpha\gamma}^\dagga\,\Delta(\bRp)_{\beta\delta}^*}
\, \langle \phi|\,c^\dagga_{\bR,\gamma}\,c^\dagger_{\bRp,\delta}\,|\phi\rangle\Bigg]\nonumber\\
&& + \sum_\bR\, \Tr\Big(P_0(\bR)\,\lambda(\bR)^\dagger\,E(\bR)\,\lambda(\bR)\Big)\nonumber \\
&& \equiv 
E_{hop} + E_{int}.\label{generic-Evar}
\eea
The last term depends only on the local properties of the uncorrelated wavefunction $|\phi\rangle$, 
specifically on the occupation probabilities $P_0(\bR)$. Therefore, for any given choice of $P_0(\bR)$, 
the optimal $|\phi\rangle$ that minimizes the variational energy is the ground state of the Hamiltonian 
\bea
\mathcal{H}_{var} &=& -\sum_{\bR\bRp}\,\sum_{\alpha\beta\gamma\delta}\,t^{\alpha\beta}_{\bR\bRp}\, 
\Bigg[\nonumber \\
&& ~~~~ \sqrt{Z(\bR)_{\alpha\gamma}^\dagga\,Z(\bRp)_{\beta\delta}^*}
\, c^\dagger_{\bR,\gamma}\,c^\dagga_{\bRp,\delta}
\nonumber\\
&& ~~~~+  
\sqrt{Z(\bR)_{\alpha\gamma}^\dagga\,\Delta(\bRp)_{\beta\delta}^*}
\, c^\dagger_{\bR,\gamma}\,c^\dagger_{\bRp,\delta}
\nonumber\\
&& ~~~~+
\sqrt{\Delta(\bR)_{\alpha\gamma}^\dagga\,Z(\bRp)_{\beta\delta}^*}
\, c^\dagga_{\bR,\gamma}\,c^\dagga_{\bRp,\delta}
\nonumber\\
&& ~~~~+
\sqrt{\Delta(\bR)_{\alpha\gamma}^\dagga\,\Delta(\bRp)_{\beta\delta}^*}
\, c^\dagga_{\bR,\gamma}\,c^\dagger_{\bRp,\delta}\Bigg] \nonumber\\
&& -\sum_{\bR}\, \sum_{\alpha\beta}\,\Bigg[ 
\mu(\bR)_{\alpha\beta}\,c^\dagger_{\bR,\alpha}c^\dagga_{\bR,\beta} \nonumber \\
&&~~~~+ \Big(\nu(\bR)_{\alpha\beta}\,c^\dagger_{\bR,\alpha}c^\dagger_{\bR,\beta} + H.c.\Big)
\Bigg],\label{var-Ham}
\eea
where the parameters $\mu(\bR)_{\alpha\beta}$ and $\nu(\bR)_{\alpha\beta}$ are Lagrange multipliers to be 
determined by imposing that the ground state has indeed the chosen $P_0(\bR)$. The last task is to 
find the values of the variational parameters $\lambda(\bR)_{\Gamma\Gamma'}$  as well as of $P_0(\bR)$ for which 
the variational energy (\ref{generic-Evar}) is minimum. We note that
the variational Hamiltonian (\ref{var-Ham}) that has to be solved may include also inter-site pairing terms, 
which are absent in the original Hamiltonian (\ref{generic-Ham}). 

Analogously to other more conventional variational approaches, like Hartree-Fock theory, it is common to interpret the single-particle 
spectrum of the variational Hamiltonian (\ref{var-Ham}) as an approximation of the true coherent spectrum of 
quasi-particles.\cite{Gebhard-FL}

\section{The model}
\label{Sec.II}

Let us now apply the variational wavefunction to specific models that are inspired by the 
valence-bond crystal example we mentioned in the introduction, and where the off-diagonal elements of the  
operator $\mathcal{P}_\bR$ as well as its non-hermitean character do play an important role. Since the operator 
$\mathcal{P}_\bR$ is built out of purely local properties, namely the available on-site electronic configurations  
plus a variational guess for the uncorrelated on-site single-particle density-matrix, a lot of preliminary results 
can be extracted without even specifying how lattice-sites are coupled together. 
Therefore we start our analysis from defining some local properties and 
later we will consider a specific lattice model.   

\subsection{The isolated dimer}

The basic unit of the model we are going to investigate consists of a dimer with Hamiltonian 
\bea
\mathcal{H}_{dimer} &=& -t_\perp\,\sum_\sigma \, \Big(c^\dagger_{1\sigma}c^\dagga_{2\sigma} + H.c.\Big) 
+ \frac{U}{2}\sum_{i=1}^2\, \left(n_{i}-1\right)^2\nonumber \\
&& \equiv \mathcal{H}_\perp + \mathcal{H}_U,
\label{Ham-dimer}
\eea
where 1 and 2 refer to the two sites of the dimer and $n_i$, $i=1,2$, is the on-site occupation number.

It is more convenient to work in the basis of the even (bonding) and odd (anti-bonding) combinations, defined through 
\[
c_{e\sigma} = \frac{1}{\sqrt{2}}\left(c_{1\sigma} + c_{2\sigma}\right),\;\;
c_{o\sigma} = \frac{1}{\sqrt{2}}\left(c_{1\sigma} - c_{2\sigma}\right).
\]
and use this basis to built the available electronic configurations, which we will denote as $|n,\Gamma\rangle$, 
with $n$ that refers to the number of electrons. The empty and the fourfold occupied dimer 
states are denoted as $|0\rangle$ and $|4\rangle$, respectively, while the singly-occupied states as 
\[
|1,e(o),\sigma\rangle = c^\dagger_{e(o)\sigma}\,|0\rangle, 
\]
and the states with 3 electrons as 
\[
|3,e(o),\sigma\rangle = c^\dagger_{e(o)\sigma}\,
c^\dagger_{o(e)\uparrow}\,c^\dagger_{o(e)\downarrow}\,|0\rangle.
\]
There are six doubly-occupied configurations. Two are spin-singlets with two electrons 
in the even or in the odd orbital, $|2,e\rangle$ and $|2,o\rangle$, respectively. When each 
orbital is singly occupied, the two electrons form either a spin triplet, $|2,1,S_z\rangle$ with $S_z=-1,0,1$, 
or a spin singlet, $|2,0\rangle$. 
Since we are not going to consider variational solutions that break spin-$SU(2)$ symmetry, it is convenient 
to define the projector operators 
\ba
\Pj{1,e(o)}{1,e(o)} &=& \sum_\sigma\, \Pj{1,e(o),\sigma}{1,e(o),\sigma},\\
\Pj{3,e(o)}{3,e(o)} &=& \sum_\sigma\, \Pj{3,e(o),\sigma}{3,e(o),\sigma},\\
\Pj{2,1}{2,1} &=& \sum_{S_z=-1}^1\, \Pj{2,1,S_z}{2,1,S_z}.
\ea

The isolated-dimer ground state in the subspace with two electrons is 
\ba
|\Psi\rangle &=& \frac{\cos\theta}{\sqrt{2}}\left(c^\dagger_{1\uparrow}c^\dagga_{2\downarrow}
+c^\dagger_{2\uparrow}c^\dagga_{1\downarrow}\right)\,|0\rangle \\
&& + \frac{\sin\theta}{\sqrt{2}}\left(c^\dagger_{1\uparrow}c^\dagga_{1\downarrow}
+c^\dagger_{2\uparrow}c^\dagga_{2\downarrow}\right)\,|0\rangle\\
&=& \frac{1}{\sqrt{2}}\left(\cos\theta + \sin\theta\right)\,|2,e\rangle \\
&& - \frac{1}{\sqrt{2}}\left(\cos\theta - \sin\theta\right)\,|2,o\rangle,
\ea
where $\tan 2\theta = 4t_\perp/U$ and has energy 
\be
E = \frac{U}{2} - \sqrt{\left(\frac{U}{2}\right)^2 + 4t_\perp^2}.\label{E-isolated}
\ee 
$|\Psi\rangle$ can be always rewritten in the form of a 
Gutzwiller wavefunction. First of all, we needs to choose an uncorrelated wavefunction $|\phi\rangle$. A natural choice 
might be the ground state at $U=0$, namely $|2,e\rangle$. Indeed 
$|\Psi\rangle$ can be written as 
\[
|\Psi\rangle = \mathcal{P}\,|2,e\rangle,
\]
where 
\bea
\mathcal{P} &=& \Pj{\Psi}{2,e} = \frac{1}{\sqrt{2}}\left(\cos\theta + \sin\theta\right)\,\Pj{2,e}{2,e}\nonumber \\ 
&& -\frac{1}{\sqrt{2}}\left(\cos\theta - \sin\theta\right)\,\Pj{2,o}{2,e}.
\label{P-isolated}
\eea
and obviously satisfies both (\ref{cond-1}) and (\ref{cond-2}). 

Another possibility, that we are also going to consider in what follows, is to use an uncorrelated wavefunction that corresponds 
to a dimer in which the two sites are only coupled by an intersite singlet-Cooper pairing, 
namely with $\langle c^\dagger_{1\uparrow}c^\dagger_{2\downarrow}\rangle 
= \langle c^\dagger_{2\uparrow}c^\dagger_{1\downarrow}\rangle \not = 0 $. In this case 
\[
|\phi\rangle = \frac{1}{2}\Big(|0\rangle + |2,e\rangle - |2,o\rangle - |4\rangle\Big),
\]
and, once again, the true ground state can be written as 
\[
|\Psi\rangle = |\Psi\rangle\langle \phi|\,|\phi\rangle \equiv \mathcal{P}\, |\phi\rangle.
\]
Already at this stage one can appreciate how important is the role 
of the off-diagonal elements in $\mathcal{P}$, especially for large $U/t_\perp$.

\subsection{The non-isolated dimer: variational density matrix}

When the dimer is coupled to the rest of the system, in order to built the operator $\mathcal{P}$ we need 
to specify an uncorrelated local single-particle density matrix based on a variational guess of the 
uncorrelated wavefunction $|\phi\rangle$. A simple guess would be a magnetic wavefunction in which 
the two sites of each dimer have opposite magnetization. This choice is also the only one admitted by 
an Hartree-Fock decomposition of the interaction term $\mathcal{H}_U$. However, a magnetic wavefunction 
is not the most suitable choice to reproduce the limit of isolated dimers, which is a collection 
of singlets. 

Alternatively, one can consider a paramagnetic $|\phi\rangle$ that has built in the tendency 
of each dimer to lock into a spin-singlet. This can be accomplished in two ways that do not exclude each other. 
The first is to assume an uncorrelated wavefunction with a huge splitting between even and odd orbitals, 
namely with 
\[
n_e = \sum_\sigma\, \langle \phi|\,c^\dagger_{e\sigma}c^\dagga_{e\sigma}\,|\phi\rangle 
\gg n_o = \sum_\sigma\, \langle \phi|\,c^\dagger_{o\sigma}c^\dagga_{o\sigma}\,|\phi\rangle.
\] 
This implies that, among the doubly-occupied configurations of each dimer, mainly the spin-singlet 
$|2,e\rangle$ survives in the uncorrelated wavefunction. The latter can then be turned into the isolated dimer 
configuration by an appropriate Gutzwiller operator $\mathcal{P}$, as shown before. 
The other possibility is to include Cooper pairing correlations in the singlet channel 
\[
\Delta_{SC} = \langle \phi |\,c^\dagger_{1\uparrow}c^\dagger_{2\downarrow}\, + 
\,c^\dagger_{2\uparrow}c^\dagger_{1\downarrow}\,|\phi\rangle.
\] 
In this case, the isolated dimer can be recovered by assuming a very strong pairing $\Delta_{SC}\simeq 1$ 
and suppressing, through $\mathcal{P}$, configurations with none or two singlet-pairs. 
Note that both $n_e-n_o$ and $\Delta_{SC}$ do not appear by a mean-field decoupling of $\mathcal{H}_U$, so 
that a variational wavefunction with such correlations built in can not be stabilized within Hartree-Fock 
theory. Here the role of $\mathcal{P}$ becomes crucial. 

Therefore, let us assume for $|\phi\rangle$ a BCS-wavefunction defined such that  
\bea
&&\langle \phi|\, c^\dagger_{1\sigma}c^\dagga_{1\sigma} \,|\phi\rangle 
= \langle \phi|\, c^\dagger_{2\sigma}c^\dagga_{2\sigma} \,|\phi\rangle = \frac{n}{4},\label{C:n}\\
&&\langle \phi|\, c^\dagger_{1\sigma}c^\dagga_{2\sigma} \,|\phi\rangle 
= \langle \phi|\, c^\dagger_{2\sigma}c^\dagga_{1\sigma} \,|\phi\rangle = \frac{\delta}{2},\label{C:delta}\\
&&\langle \phi|\, c^\dagger_{1\uparrow}c^\dagger_{2\downarrow} \,|\phi\rangle 
= \langle \phi|\, c^\dagger_{2\uparrow}c^\dagger_{1\downarrow} \,|\phi\rangle 
=\frac{\Delta_{SC}}{2},\label{C:Delta}
\eea
with real $\Delta_{SC}$. In the even/odd basis this translates into 
\bea
&&\langle \phi|\, c^\dagger_{e\sigma}c^\dagga_{e\sigma} \,|\phi\rangle = \frac{n_e}{2},\label{C:ne}\\ 
&&\langle \phi|\, c^\dagger_{o\sigma}c^\dagga_{o\sigma} \,|\phi\rangle = \frac{n_o}{2},\label{C:no}\\ 
&&\langle \phi|\, c^\dagger_{e\uparrow}c^\dagger_{e\downarrow} \,|\phi\rangle 
= - \langle \phi|\, c^\dagger_{o\uparrow}c^\dagger_{o\downarrow} \,|\phi\rangle 
= \frac{\Delta_{SC}}{2}\label{C:Deltaeo},
\eea
where $n_e+n_o=n$. As previously mentioned, the calculations simplify 
considerably in the natural basis, that is derived 
in the Appendix for this particular choice of density matrices. 

As a particular application, we assume hereafter that the model is half-filled, namely $n_e+n_0=2$. 
The density matrix of the operators in the natural basis, $d^\dagga_{e(o)\sigma}$ 
and $d^\dagger_{e(o)\sigma}$ is, by Eqs.~(\ref{APP:C-natural}) and (\ref{APP:q}),      
\[
\langle \phi|\, d^\dagger_{e(o)\sigma} d^\dagga_{e(o)\sigma}\,|\phi\rangle = 
\frac{1}{2} + q,
\]
where 
\be
q = \frac{1}{2}\,\sqrt{\delta^2 + \Delta_{SC}^2}.
\label{def:q}
\ee
The two angles $\theta_e$ and $\theta_o$, that are defined by Eq.~(\ref{APP:theta})
and identify the unitary transformation from the original to the natural basis, are given by  
$\theta_e = \theta$ and $\theta_o=\theta-\pi/2$, where 
\be
\tan 2\theta = \frac{\Delta_{SC}}{\delta}.
\label{def:theta}
\ee
We note that, for $q\to 1/2$, the uncorrelated wavefunction describes an insulator where charge fluctuations 
are completely suppressed since each natural orbital is fully occupied. It is obvious that, if our choice of the 
variational wavefunction is correct, then {\bf the optimal uncorrelated wavefunction must asymptotically acquire 
$\mathbf{q=1/2}$ for $\mathbf{U\to\infty}$}. 

The expression in the natural basis of the hopping, Eq.~(\ref{APP:hopping}), 
and interaction, Eq.~(\ref{APP:U}), operators 
can be derived through (\ref{APP:transform}) and have a relatively simple expression  
at half-filling:
\bea
\mathcal{H}_\perp &=& -t_\perp\,\Bigg[2\cos 2\theta\,\Big(\Pj{\wt{4}}{\wt{4}} - \Pj{\wt{0}}{\wt{0}}\Big)\label{hopping-natural}\\
&& + \cos 2\theta \,\Big(\Pj{\wt{3}}{\wt{3}} - \Pj{\wt{1}}{\wt{1}}\Big) - 
\sin 2\theta \,\Big(\Pj{\wt{1}}{\wt{3}} + H.c.\Big)\nonumber\\
&& - \sqrt{2}\,\sin 2\theta \, \Big(\Pj{\wt{0}}{\wt{2},+} 
+ \Pj{\wt{4}}{\wt{2},+} + H.c.\Big)\Bigg]\nonumber \\
\mathcal{H}_U &=& \frac{U}{2} \,\Big(\Pj{\wt{0}}{\wt{0}}
+ \Pj{\wt{4}}{\wt{4}} - \Pj{\wt{0}}{\wt{4}} - \Pj{\wt{4}}{\wt{0}}\Big)\nonumber \\
&& + U\, \Big(\Pj{\wt{2},+}{\wt{2},+} + 
\Pj{\wt{2},-}{\wt{2},-} + \Pj{\wt{2},0}{\wt{2},0}\Big)\nonumber \\
&& + \frac{U}{2}\,\Big(\Pj{\wt{1}}{\wt{1}}+\Pj{\wt{3}}{\wt{3}}\Big)\label{natural-U},
\eea
where we have defined 
\ba
\Pj{\wt{1}(\wt{3})}{\wt{1}(\wt{3})} &=& \Pj{\wt{1}(\wt{3}),e}{\wt{1}(\wt{3}),e} + 
\Pj{\wt{1}(\wt{3}),o}{\wt{1}(\wt{3}),o},\\
\Pj{\wt{1}}{\wt{3}} &=& \Pj{\wt{1},e}{\wt{3},e} + \Pj{\wt{1},o}{\wt{3},o},\\
|\wt{2},\pm\rangle &=& \frac{1}{\sqrt{2}}\,\Big(|\wt{2},e\rangle \pm |\wt{2},o\rangle\Big),
\ea
and denoted the local configurations in the natural basis as $|\wt{n},\Gamma\rangle$ to distinguish them from 
the analogous ones in the original representation.

\subsection{The Gutzwiller operator $\mathcal{P}$}

The most general Gutzwiller operator $\mathcal{P}$ should include at least all the projectors  
$\Pj{\wt{n},\Gamma}{\wt{n},\Gamma}$ as well as all the off-diagonal operators 
$\Pj{\wt{n},\Gamma}{\wt{n'},\Gamma'}$ that appear in the local Hamiltonian, Eqs.~(\ref{hopping-natural}) 
and (\ref{natural-U}). As we mentioned before, our expectation is that the uncorrelated wavefunction which  
better connects to the large-$U$ Mott insulator should be identified by $q\to 1/2$, in which locally only 
the configurations $|\wt{3}\rangle$ and $|\wt{4}\rangle$ are occupied with non-negligible probability. 
This suggests that $\mathcal{P}$ must include at least 
those off-diagonal operators that would turn $|\wt{4}\rangle$ into the isolated dimer ground state, namely 
$\Pj{\wt{0}}{\wt{4}}$ and $\Pj{\wt{2},+}{\wt{4}}$. The latter forces to include also $\Pj{\wt{1}}{\wt{3}}$, 
as we shall see.

Therefore we assume for $\mathcal{P}$ the following variational ansatz:
\bea
\mathcal{P} &=& \sum_{\wt{n}\Gamma}\,\lambda_{n\Gamma}\,\Pj{\wt{n},\Gamma}{\wt{n},\Gamma} + 
\lambda_{13}\, \Pj{\wt{1}}{\wt{3}}\nonumber \\
&& + \lambda_{04}\,\Pj{\wt{0}}{\wt{4}} + \lambda_{2+\,4}\,\Pj{\wt{2},+}{\wt{4}},
\label{ansatz-P}
\eea
with real $\lambda$'s. We define 
\be
P(n,\Gamma) = \lambda_{n\Gamma}^2\, P_0(\wt{n},\Gamma),\label{P:n not=34}
\ee
for all $n\not = 3,4$, while, for $n=3,4$,   
\bea
P(3) &=& \left(\lambda_3^2 + \lambda_{13}^2\right)\, P_0(\wt{3}),\label{P:n=3}\\
P(4) &=& \left(\lambda_4^2 + \lambda_{04}^2 + \lambda_{2+\,4}^2\right)\,P_0(\wt{4}).\label{P:n=4}
\eea
Then the conditions Eqs.~(\ref{cond-1}) and (\ref{cond-2}) read 
\bea
&& \sum_{n\Gamma}\, P(n,\Gamma) = 1,\label{uno}\\
&& \sum_{n\Gamma}\, n\,P(n,\Gamma) = 2+4q,\label{due}\\
&&\lambda_{13}\,\lambda_1\,\sqrt{P_0(\wt{3})\,P_0(\wt{1})} = \nonumber \\ 
&&~~- \sqrt{2}\,\lambda_{2+\,4}\,\lambda_{2+}\,\sqrt{P_0(\wt{4})\,P_0(\wt{2},+)}.\label{tre}
\eea
Here $P_0(\wt{n},\Gamma)$ are the occupation probabilities in the natural basis of the uncorrelated wavefunction. 
Specifically
\[
P_0(\wt{n},\Gamma) = g_{\wt{n},\Gamma}\, \left(\frac{1}{2}+q\right)^{\wt{n}}\; \left(\frac{1}{2}-q\right)^{4-\wt{n}},
\]
where $g_{\wt{n},\Gamma}$ is the degeneracy of the configuration. 
Eq.~(\ref{tre}) guarantees that the anomalous averages 
\[
\langle \phi|\,\mathcal{P}^\dagger\,\mathcal{P}\, d^\dagger_{e(o)\uparrow}d^\dagger_{e(o)\downarrow}\,|\phi\rangle
\]
vanish in the natural basis, and explains why we have included $\Pj{\wt{1}}{\wt{3}}$ in (\ref{ansatz-P}). 
It is convenient to rewrite  
\ba
\lambda_i &=& \sqrt{\frac{P(4)}{P_0(\wt{4})}}\, u_i \qquad \mbox{for}~i=4,~04,~2+4,\\
\lambda_i &=& \sqrt{\frac{P(3)}{P_0(\wt{3})}}\, u_i \qquad \mbox{for}~i=3,~13,
\ea
where $u_3^2 + u_{13}^2 = 1$, which can be satisfied by choosing  
$u_3=\cos\psi$ and $u_{13}=\sin\psi$, and $u_4^2 + u_{04}^2 + u_{2+\,4}^2 =1$. The latter parameters 
can be expressed by means of another unit vector $\mathbf{v}=(v_1,v_2,v_3)$, through 
\be
\begin{array}{lcl}
v_1 &=& \frac{1}{\sqrt{2}}\,\left(u_4 + u_{04}\right),\\
v_2 &=& \frac{\cos 2\theta}{\sqrt{2}}\,\left(u_4 - u_{04}\right)
- \sin 2\theta\,u_{2+\,4},\\
v_3 &=& \frac{\sin 2\theta}{\sqrt{2}}\,\left(u_4 - u_{04}\right)
+ \cos 2\theta\,u_{2+\,4},\\
\end{array}
\label{v-u}
\ee
In terms of all the variational parameters, the $P(n,\Gamma)$'s, $\theta$, $q$, $\psi$ and $\mathbf{v}$,
the average values per dimer of the interaction, $\mathcal{H}_U$, and intra-dimer hopping, $\mathcal{H}_\perp$, 
are readily found to be  
\bea
E_U &=& \langle \phi|\,\mathcal{P}^\dagger\,\mathcal{H}_U\,\mathcal{P}\,|\phi\rangle 
= \frac{U}{2}\,\Big(P(3)+P(1)\Big)\nonumber \\
&& + \frac{U}{2}\,P(0) + U\,\Big(P(2,+)+P(2,-)+P(2,0)\Big)\nonumber \\
&& +  U\Big(v_2^2+v_3^2\Big)\,P(4),\label{av-U} \\
E_\perp &=& \langle \phi|\,\mathcal{P}^\dagger\,\mathcal{H}_\perp\,\mathcal{P}\,|\phi\rangle = 
- 2\,t_\perp\,\delta_*,
\label{av-tperp}
\eea
where the actual correlated values of the hybridization and of the anomalous average are 
\bea
2\,\delta_* &=& \langle \phi|\, \mathcal{P}\,\left(n_e - n_o\right)\,\mathcal{P}\,|\phi\rangle \nonumber \\
&=& 4\,v_1\,v_2\,P(4)+ \cos \left(2\theta + 2\psi\right)\,P(3) \nonumber \\
&&  -2\cos 2\theta\,P(0) - \cos 2\theta\,P(1),
\label{delta*}\\
2\Delta_* &=& \langle \phi|\, \mathcal{P}\,\left(c^\dagger_{e\uparrow}c^\dagger_{e\downarrow}
- c^\dagger_{o\uparrow}c^\dagger_{o\downarrow} + H.c.\right)\, \mathcal{P}\,|\phi\rangle\nonumber \\
&=& 4\,v_1\,v_3\, P(4) + \sin \left(2\theta + 2\psi\right)\,P(3) \nonumber \\
&&  -2\sin 2\theta\,P(0) -\sin 2\theta \,P(1),\label{Delta*}\\
\eea
We note that $\delta_*$ and $\Delta_*$ are mutually exclusive, namely the choice of parameters that maximizes one 
of the two, makes the other vanishing.

Upon the action of $\mathcal{P}$, the single fermion operators in the Nambu spinor representation 
transform effectively into 
\be
\mathcal{P}^\dagger\,\left(
\begin{array}{c}
d_{e(o)\uparrow}^\dagga\\
d_{e(o)\downarrow}^\dagger \\
\end{array}
\right)\,\mathcal{P} \rightarrow
\sqrt{Z+\Delta}\;\; {\rm e}^{-i\,\beta\,\tau_2\,}
\, \left(
\begin{array}{c}
d_{e(o)\uparrow}^\dagga\\
d_{e(o)\downarrow}^\dagger \\
\end{array}
\right),
\label{Project-phi}
\ee
where $\tau_i$, $i=1,2,3$, are the Pauli matrices that act on the Nambu spinor components, 
\be
\tan \beta = \frac{\sqrt{\Delta}}{\sqrt{Z}}.\label{beta} 
\ee
and, finally, 
\bea
\sqrt{Z} &=& \frac{\displaystyle \langle \phi|\,\mathcal{P}^\dagger\,d_{e(o)\sigma}^\dagger\,\mathcal{P}\,
d_{e(o)\sigma}^\dagga\,|\phi\rangle}{\displaystyle \frac{1}{2}+q}\nonumber \\
&=& \sqrt{\frac{1}{1-4q^2}}\,\Bigg[
\sqrt{P(0)\,P(1)} + \frac{1}{2}\sqrt{P(1)\,P(2,+)}\nonumber \\
&& + \frac{1}{2}\sqrt{P(1)\,P(2,-)} 
+ \frac{1}{2}\sqrt{P(1)\,P(2,0)} \nonumber \\
&& + \frac{\sqrt{3}}{2}\sqrt{P(1)\,P(2,1)}
+ \cos\psi\,\,\frac{\sqrt{3}}{2}\sqrt{P(3)\,P(2,1)}\nonumber \\
&& + \cos\psi\,\frac{1}{2}\sqrt{P(3)\,P(2,0)}
+ \cos\psi\,\frac{1}{2}\sqrt{P(3)\,P(2,-)}\nonumber \\
&& + \cos\psi\,\frac{1}{2}\sqrt{P(3)\,P(2,+)}\nonumber\\
&& + \frac{1}{\sqrt{2}}\,\Big(v_1\,\cos\psi + v_2\,\cos\left(2\theta+\psi\right) \nonumber \\
&&~~~+ v_3\,\sin\left(2\theta+\psi\right)\Big)
\,\sqrt{P(4)\,P(3)}\Bigg],\label{sqrt_Z}\\
\sqrt{\Delta} &=& \frac{\displaystyle \langle \phi|\,\mathcal{P}^\dagger\,d_{e(o)\uparrow}^\dagger\,\mathcal{P}\,
d_{e(o)\downarrow}^\dagger\,|\phi\rangle}{\displaystyle \frac{1}{2}-q}\nonumber \\
&=& \sqrt{\frac{1}{1-4q^2}}\,\Bigg[
\frac{1}{2}\,\sin\psi\,\sqrt{P(3)\,P(2,-)}\nonumber \\
&& -\frac{1}{2}\,\sin\psi\,\sqrt{P(3)\,P(2,+)}\nonumber\\
&& + \frac{\sqrt{3}}{2}\,\sin\psi\,\sqrt{P(3)\,P(2,1)}\nonumber \\
&& + \frac{1}{2}\,\sin\psi\,\sqrt{P(3)\,P(2,0)}\nonumber \\
&& +\frac{1}{\sqrt{2}}\,\Big(v_1\,\sin\psi - v_2\,\sin\left(2\theta+\psi\right)\nonumber \\
&&  + v_3\,\cos\left(2\theta+\psi\right)\Big)
\,\sqrt{P(4)\,P(3)}\Bigg], \label{sqrt_Delta}
\eea
with real $\sqrt{Z}$ and $\sqrt{\Delta}$. Therefore, if the dimers are coupled one to another by the single particle 
hopping term
\be
\mathcal{T} = \sum_{\bR\not= \bRp}\,\sum_{i,j=e,o}\,
t^{ij}_{\bR\bRp}\, \Psi^\dagger_{\bR,i}\,\tau_3\, \Psi^\dagga_{\bRp,j},
\label{dimer-hopping-T}
\ee
where 
\[
\Psi^\dagger_{\bR,i} = (c^\dagger_{\bR,i\uparrow},c^\dagga_{\bR,i\downarrow}),
\]
and $\Psi$ its hermitean conjugate, the uncorrelated wave function $|\phi\rangle$ minimizes the effective hopping
\be 
\mathcal{T}_{var} = \left(Z+\Delta\right)\,\sum_{\bR\not= \bRp}\,\sum_{i,j=e,o}\,
t^{ij}_{\bR\bRp}\, \Psi^\dagger_{\bR,i}\,\tau_3\, {\rm e}^{-2\,i\,\beta\,\tau_2}\,\Psi^\dagga_{\bRp,j},
\label{dimer-hopping-T-var}
\ee 
under the condition that the local density matrix satisfies Eqs.(\ref{C:ne})-(\ref{C:Deltaeo}). 
One can readily show that this amounts to find the ground state $|\phi\rangle$ of the variational Hamiltonian
\bea
\mathcal{H}_{var} &=& \left(Z+\Delta\right)\,\sum_{\bR\not= \bRp}\,\sum_{i,j=e,o}\,
t^{ij}_{\bR\bRp}\, \Psi^\dagger_{\bR,i}\,\tau_3\, \Psi^\dagga_{\bRp,j} \nonumber\\
&-& \sum_\bR\, \mu_3\,\Big(\Psi^\dagger_{\bR,e}\,\tau_3\, \Psi^\dagga_{\bR,e}
- \Psi^\dagger_{\bR,o}\,\tau_3\, \Psi^\dagga_{\bR,o}\Big)\nonumber\\
&-& \sum_\bR\, \mu_1\,\Big(\Psi^\dagger_{\bR,e}\,\tau_1\, \Psi^\dagga_{\bR,e}
- \Psi^\dagger_{\bR,o}\,\tau_1\, \Psi^\dagga_{\bR,o}\Big),\label{Dimer-Ham-var}
\eea
with $\mu_3$ and $\mu_1$ such that 
\[
\langle \phi|\, \mathcal{H}_{var}\,|\phi\rangle + 4q\,\mu_3\,\cos(2\theta+2\beta) + 
4q\,\mu_1\,\sin(2\theta+2\beta),
\]
is maximum.

\medskip

Before we consider specific lattice models, it is worth re-deriving within this variational scheme 
the isolated-dimer ground-state energy (\ref{E-isolated}) at half-filling. For that purpose, we take all $P(n,\Gamma)$ zero 
but $P(4)=1$. The variational energy is simply  
\[
E_{var} = E_U + E_\perp =  U\,\left(v_2^2 + v_3^3\right) -4\,t_\perp\,v_1\,v_2.
\]
The minimum under the constraint $\mathbf{v}\cdot\mathbf{v} = 1$ is obtained for $v_3=0$ and 
exactly reproduces (\ref{E-isolated}). We note that the minimum energy is independent on $\theta$, 
namely there exists a continuous family of variational solutions with equal energy parametrized by $\theta$. 
However, in spite of the fact that the uncorrelated wavefunction may describe a superconductor, the actual value of 
the anomalous average $\Delta_* =0$.   

\section{A lattice model of dimers}
\label{Sec.III}

As a particular application, let us consider the following lattice model
\bea
\mathcal{H} &=& -\,\sum_{\bR\bRp}\,\sum_{i=1}^2\,\sum_\sigma \, t_{\bR\bRp}\, 
c^\dagger_{\bR,i\sigma}c^\dagga_{\bRp,i\sigma} + H.c.\nonumber\\
&& + \frac{U}{2}\sum_{\bR}\,\sum_{i=1}^2\,\left(n_{\bR,i}-1\right)^2\nonumber\\
&& -t_\perp\,\sum_{<\bR\bRp>}\,\sum_\sigma\,c^\dagger_{\bR,1\sigma}c^\dagga_{\bR,2\sigma} + H.c.\nonumber\\
&=& \sum_{\bk\sigma}\,\left(\epsilon_{\bk}-t_\perp\right)\,c^\dagger_{\bk,e\sigma}c^\dagga_{\bk,e\sigma}
+ \left(\epsilon_{\bk}+t_\perp\right)\,c^\dagger_{\bk,o\sigma}c^\dagga_{\bk,o\sigma} \nonumber\\ 
&& + \frac{U}{2}\sum_{\bR}\,\sum_{i=1}^2\,\left(n_{\bR,i}-1\right)^2,
\label{Ham-2planes}
\eea
where $n_{\bR,i} = \sum_\sigma \, c^\dagger_{\bR,i\sigma}c^\dagga_{\bR,i\sigma}$ and $\epsilon_\bk$ is the band dispersion 
induced by $t_{\bR\bRp}$, with half-bandwidth $D$.
The Hamiltonian (\ref{Ham-2planes}) represents two 
Hubbard models coupled by a single-particle hopping $t_\perp$, each model being defined on a lattice with 
coordination number $z$. As we mentioned, the variational results that we have so far derived are rigorous strictly 
speaking only if $z\to\infty$, 
although, in the spirit of the Gutzwiller approximation, they can be used for generic $z$ as well. 

If $U\gg D,t_\perp$, (\ref{Ham-2planes}) describes at half-filling a Mott insulator which may be magnetic at $t_\perp\ll D$, but is 
certainly non-magnetic at $t_\perp\gg D$, where the ground state reduces essentially to a collection of singlets. 
For instance, in the case of 
a Bethe lattice with nearest neighbor hopping, the transition is at $t_\perp = D/\sqrt{8}$, 
value that is going to decrease if frustration is 
included. If $U$ is small and the Fermi surface is not nested, then the model is metallic for $t_\perp\leq D$ 
and is a band insulator otherwise. In fact, in the absence of nesting there is 
generically a finite window of $t_\perp$ values in which, 
upon increasing $U$, the model undergoes a transition from a paramagnetic metal into a non-magnetic Mott insulator, 
and this is just the case we are going to consider in what follows. 
The same model have been recently studied by Fuhrmann, Heilmann and Monien using DMFT~\cite{Monien} 
and by Kancharla and Okamoto~\cite{Okamoto} using DMFT and cluster DMFT, respectively,  
that gives us the opportunity to directly check the accuracy of our wavefunction.

The variational Hamiltonian (\ref{Dimer-Ham-var}) of the model (\ref{Ham-2planes}) has a very simple expression, 
\bea
\mathcal{H}_{var} &=& \sum_{\bk}\,\Psi^\dagger_{\bk,e}\,
\big[\big(\epsilon_{\bk*} - \mu_3\big)\,\tau_3 - \mu_1\,\tau_1\Big]\, \Psi^\dagga_{\bk,e}\nonumber\\
&& + \Psi^\dagger_{\bk,o}\,
\big[\big(\epsilon_{\bk*} + \mu_3\big)\,\tau_3 + \mu_1\,\tau_1\Big]\, \Psi^\dagga_{\bk,o},\label{Ham-var-2planes}
\eea
where 
\[
\epsilon_{\bk*} = \left(Z+\Delta\right)\,\epsilon_\bk.
\]
The variational single-particle spectrum has the conventional BCS form with eigenvalues 
\[
E_{e\bk} = \sqrt{\left(\epsilon_{\bk*}-\mu_3\right)^2 + \mu_1^2},\;
E_{o\bk} = \sqrt{\left(\epsilon_{\bk*}+\mu_3\right)^2 + \mu_1^2},
\]
hence, for any $\mu_1\not = 0$, has a gap equal to $2\mu_1$. 
On the contrary, when $\mu_1=0$, the spectrum is gapless for $|\mu_3|\leq D$, otherwise is gaped. 
The Lagrange multipliers $\mu_1$ and $\mu_2$ are obtaining by maximizing
\bea
&& E_{hop} = -\sum_\bk\,\big(E_{e\bk} + E_{o\bk}\big)
\nonumber \\
&&~~ +  4q\,\mu_3\,\cos(2\theta+2\beta) \,+\, 
4q\,\mu_1\,\sin(2\theta+2\beta)\label{E_hop-2planes}.
\eea
In terms of (\ref{E_hop-2planes}), (\ref{av-U}) and (\ref{av-tperp}) the variational energy is 
\be
E_{var} = E_{hop} + E_U + E_\perp,\label{var-E-2planes}
\ee
and depends on eight independent variational parameters. 

\begin{figure}
\includegraphics[width=8cm]{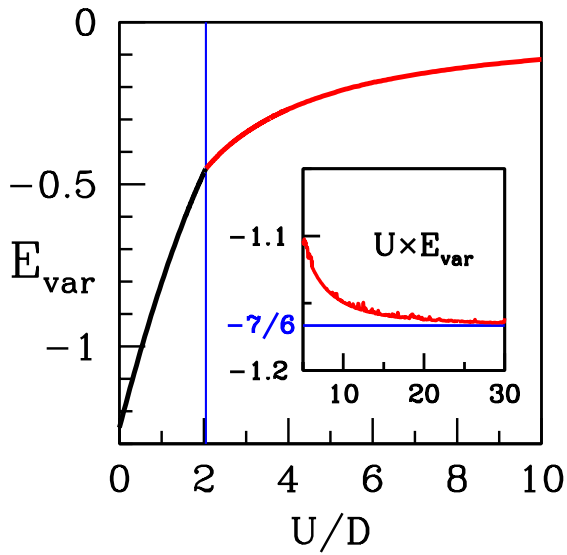}
\caption{Variational energy $E_{var}$ in units of $D$ and for a flat non-interacting density of states, 
as function of $U/D$ for $t_\perp=D/2$. At $U_c\simeq 2.05~D$ a first order 
transition occurs. The inset shows the asymptotic value of $UE_{var}/D^2$.\label{Energia}}
\end{figure}

We have solved numerically the variational problem at fixed $t_\perp/D = 0.5$ as function of $U/D$. To simplify calculations, 
we have assumed for the band dispersion $\epsilon_\bk$ either a flat or a semi-circular density of states, although both 
would give rise to nesting that could stabilize magnetic phases, which we do not take into account. However, from the point of view 
of the paramagnetic-metal to paramagnetic-insulator transition, this choice is not influential. 

\begin{figure}
\includegraphics[width=8cm]{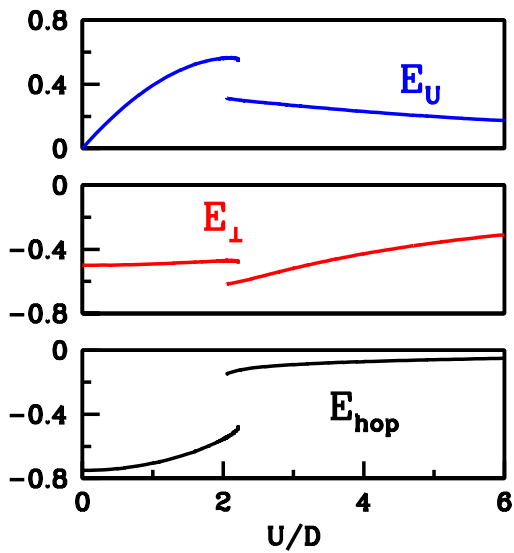}
\caption{The different contributions to the variational energy, $E_U$, $E_\perp$ and $E_{hop}$.
\label{All-energies}}
\end{figure}

We find that the variational solution 
displays a first order phase transition at $U_c\simeq 2.05~D$ for a flat density of states, as shown by the behavior of the 
variational energy in Fig.~\ref{Energia}. This result agrees almost quantitatively with the DMFT calculation~\cite{Monien} 
obtained with a semicircular density of states, that also predicts 
a first order transition with a coexistence region between $U\simeq 1.5~D$ and $1.8~D$ at the same value of $t_\perp=0.5~D$.  
We note that the energy is everywhere finite and vanishes like $1/U$ for large $U$, see 
the inset of Fig.~\ref{Energia}. The asymptotic behavior $U E_{var}/D^2 \sim - 7/6$ is compatible with second order 
perturbation theory in $t$ and $t_\perp$ using as zeroth-order state a collection of dimers, as explained below. 
In Fig.~\ref{All-energies} we show the behavior across the transition of the three contribution to the energy, 
namely $E_U$, $E_\perp$ and $E_{hop}$. We find that the transition is accompanied by an energy loss in $E_{hop}$, 
but a gain in both $E_\perp$ and $E_U$.

\begin{figure}
\includegraphics[width=8cm]{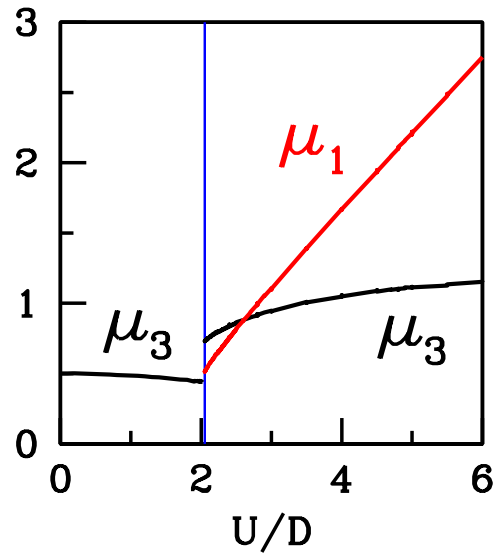}
\caption{The behavior of the parameters $\mu_1$ and $\mu_3$ in units of $D$ as function of $U/D$, 
see Eq.~(\ref{Ham-var-2planes}).
\label{mu}}
\end{figure}

\begin{figure}[t]
\includegraphics[width=8cm,height=16cm]{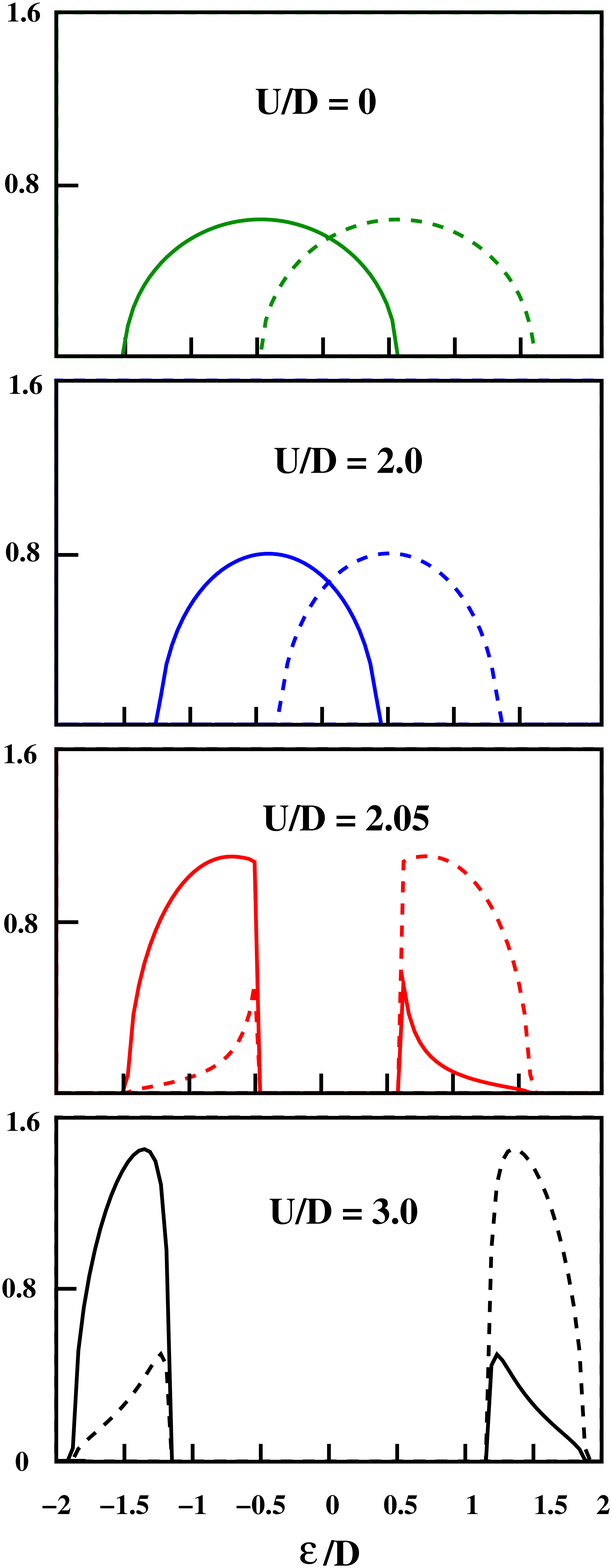}
\caption{The variational single-particle spectrum for the even, i.e. bonding, band, solid lines, and odd, i.e. 
anti-bonding, one, dashed lines, across the transition 
for a non-interacting semi-circular density of states.   
\label{DOS}}
\end{figure}

In order to characterize physically the two phases, in Fig.~\ref{mu} we plot the values 
of $\mu_1$ and $\mu_3$ across 
the transition. Since $\mu_1 = 0$, within our numerical accuracy, and $|\mu_3|<D$, the phase at $U<U_c$ is gapless hence metallic, 
see the behavior of the density of states (DOS) drawn in Fig.~\ref{DOS}. On the contrary, on the $U>U_c$ side of the transition, 
$\mu_1\not = 0$, that implies a finite gap in the single-particle variational spectrum, see Fig.~\ref{DOS}. 
In the gaped phase at $U>U_c$ the spectrum looks like the one of a Peierls insulator with 
a very large hybridization gap, not consistent with the bare value of $t_\perp$. In reality this gap is, more properly, 
the Mott-Hubbard gap. Indeed the DOS has weight both below and above the chemical potential, suggestive 
of asymmetric Mott-Hubbard side-bands. Moreover, as we are going 
to discuss below,  the actual difference between the occupations of the bonding and anti-bonding bands, 
which we denoted as $2\delta_*$ in Eq.~(\ref{delta*}), decreases with $U$, unlike the single-particle gap, see Fig.~\ref{delta}. 
This behavior is reminiscent of what has been found by Biermann 
{\sl et al.}~\cite{biermann:dimer} as an attempt to understand the physics of VO$_2$.  

\begin{figure}
\includegraphics[width=8cm]{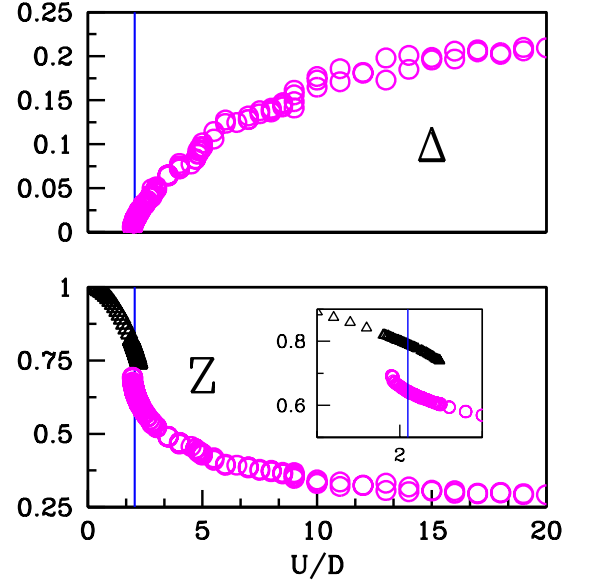}
\caption{The behavior of the parameters $Z$ and $\Delta$.
\label{Z-fig}}
\end{figure}

\begin{figure}
\includegraphics[width=8cm]{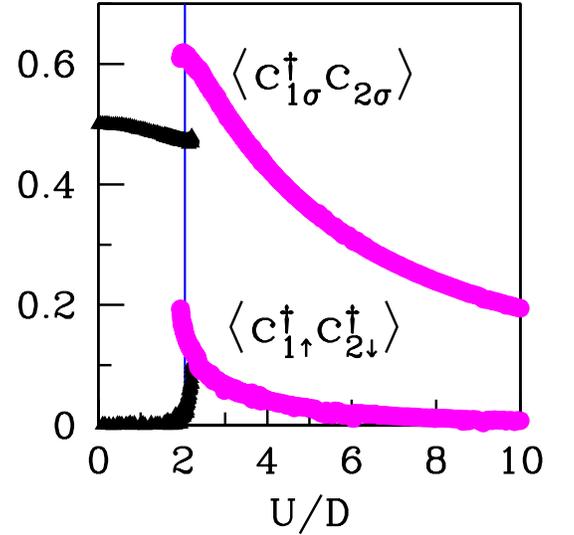}
\caption{The average values of the intra-dimer hopping, $\delta_*$, and pairing, $\Delta_*$. 
\label{delta}}
\end{figure}

The other quantities that identify the variational spectrum are $Z$ and $\Delta$, shown 
in Fig.~\ref{Z-fig}. We see that $Z$ is decreasing with $U$ but reaches a finite value $Z= 1/4$ 
for $U\to \infty$. On the contrary, $\Delta=0$ for $U<U_c$, while $\Delta \not = 0$ for $U>U_c$ 
and increases monotonically to reach asymptotically the same value 1/4 for large $U$. 

\begin{figure}
\includegraphics[width=8cm]{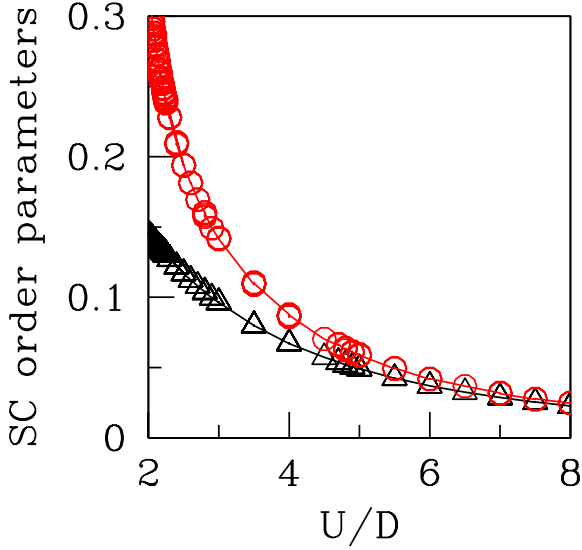}
\caption{The intra-dimer, circles, and, with reversed sign, the inter-dimer, triangles, superconducting order parameters.   
\label{SC}}
\end{figure}

Further insights can be gained by the average values of the intra-dimer hopping and pairing, 
Eqs.~(\ref{delta*}) and (\ref{Delta*}), drawn in Fig.~\ref{delta}. As we mentioned, the intra-dimer 
hybridization $2\delta_*$ is monotonically decreasing with $U$, apart from the jump at the first order transition.   
More interestingly, around the transition the variational solution 
is characterized by a sizeable BCS order parameter $\Delta_*$, which does not follow the behavior of the 
BCS coupling $\mu_1$ present in the variational Hamiltonian. Indeed, while $\mu_1$ is zero within our numerical accuracy 
for $U<U_c$, yet a non negligible $\Delta_*$ develops just before the transition, see Fig.~\ref{delta}. 
Moreover, although  $\mu_1$ starts already large for $U>U_c$ and increases monotonically with $U$, see Fig.~\ref{mu},    
the actual order parameter $\Delta_*$ is appreciable only near the transition and fastly decreases with $U$ to 
very tiny values. 
Also interesting is that, besides the intra-dimer superconducting order parameter, also an inter-dimer 
one arises, that can be for instance defined through 
\ba
\wt{\Delta}_* &=& \frac{1}{4}\,\frac{\displaystyle \sum_{i=1}^2\,\sum_{\bRp}\, t_{\bR\bRp}\,\langle \phi|
\,\mathcal{P}^\dagger\,\Psi^\dagger_{\bR,i}\,\tau_1\,\Psi^\dagga_{\bRp,i}\,\mathcal{P}\,|\phi\rangle}
{\displaystyle \sum_{\bRp}\, t_{\bR\bRp}}\\
&=& \frac{1}{4\epsilon_{\bk =\mathbf{0}}}\, \mu_1\,\frac{1}{V}\sum_\bk\, \frac{\epsilon_\bk}{E_\bk},
\ea
where $V$ is the number of sites. We find that $\wt{\Delta}_*$ has actually the opposite sign of $\Delta_*$, 
and both closely follow each other, rapidly decreasing with $U$, see 
Fig.~\ref{SC}.
Even though $\Delta_*$ and $\wt{\Delta}_*$ are everywhere finite for $U>U_c$,  
suggestive of a superconducting  phase that survives up to very large $U$, 
we believe that superconductivity, it it occurs at all, may appear 
only very close to the transition, where the value of the order parameter is larger.   
Indeed, the optimal solution with finite $\Delta_*$ and $\wt{\Delta}_*$ and another solution in which 
both are forced to be zero are practically degenerate within our numerical accuracy for large $U$.
Moreover, since the variational values of the order parameter are extremely small but close to $U_c$, the inclusion 
of quantum fluctuations, for instance in the form of a Jastrow factor as in Eq.~(\ref{GWF-Manuela}), 
would likely suppress superconductivity leading to a bona fide insulating wavefunction.     

Unfortunately, we can not compare this result with the DMFT analyses of Refs.~\onlinecite{Monien,Okamoto}, 
where superconductivity has not been looked for.~\cite{Schiro'} 

\subsection{Large $U$ limit}

In order to appreciate qualities and also single out defects of the variational wavefunction, it is worth discussing the large-$U$ 
solution. To leading order in $1/U$, one can assume all $P(n,\Gamma)=0$ but $P(3)$ and $P(4)$. The two constraints Eqs.~(\ref{uno}) 
and (\ref{due}) can be solved by defining
\[
P(3) = 2 - 4q \equiv 4d,\qquad P(4)=1-4d,
\]
with $d\ll 1$, namely $q\to 1/2$, while Eq.~(\ref{tre}) is already satisfied since $P(1)=P(2,+)=0$. 
Moreover, one readily recognizes that the variational solution 
asymptotically tends to acquire $\psi\simeq \beta \to \pi/4$, $\theta\to 0$ and $v_1\gg v_2,v_3$, 
which is indeed what we find by numerical minimization. It then follows that 
\[
\sqrt{Z} \simeq \sqrt{\Delta}\to \frac{1}{2}\,\sqrt{\frac{1}{1-4q^2}} \;\sqrt{4d(1-4d)}\to \frac{1}{2}.
\]
This implies that $\mu_3\to 0$, hence that $\mu_1$ is determined by maximizing
\[
E_{hop}  = -\frac{2}{V}\sum_\bk \sqrt{\epsilon_{\bk*}^2 +\mu_1^2} + 4q\,\mu_1,
\]
which leads to $\mu_1 = \sqrt{\epsilon_2/4d}$ and 
\be
E_{hop} = -4\,\sqrt{\epsilon_2\,d},
\label{Ehop-large-U}
\ee
where 
\[
\epsilon_2 = \frac{1}{V}\sum_\bk\,\epsilon_{\bk*}^2 = \frac{1}{4V}\sum_\bk\,\epsilon_\bk^2.
\]
At leading order, the variational energy per dimer is therefore
\[
E_{var} = -4\,\sqrt{\epsilon_2\,d} -4\,t_\perp\,v_1\,v_2 + U\,v_2^2 +2\, U\, d,
\]
that is minimized by $d=\epsilon_2/U^2$, $v_1\simeq 1$ and $v_2 \simeq  2t_\perp/U$, and takes the value 
\bea
E_{var} &=& - \frac{1}{2V}\sum_\bk\,\frac{\epsilon_\bk^2}{U} - \frac{4t_\perp^2}{U}\nonumber \\
&=& -\frac{1}{2V}\sum_{\bR\bRp}\, \frac{t_{\bR\bRp}\;t_{\bRp\bR}}{U} - \frac{4t_\perp^2}{U}.
\label{Evar-large-U}
\eea
In the case of a flat density of states 
\[
\frac{1}{V}\sum_\bk\, \delta\left(\epsilon-\epsilon_\bk\right) = \frac{1}{2D}\,\theta\left(D-|\epsilon|\right),
\]
and with $t_\perp=0.5~D$ we recover the numerical result $E_{var} \simeq -7/6\;D^2/U$, see Fig.~\ref{Energia}. 

We note that, in spite of the hybridization $\delta_* \sim t_\perp/U$ being small, the single-particle 
gap of the variational spectrum $2\mu_1 \simeq U$ is large, as one should expect in a Mott insulator.

\medskip
  
Coming back to the large-$U$ value of the variational energy (\ref{Evar-large-U}), one can readily realize that it 
coincides with the second order correction in $t_{\bR\bRp}$ to the energy of the state
\[
|\Psi\rangle =  \prod_\bR\;\sqrt{\frac{1}{2}}\,\big(
c^\dagger_{\bR,1\uparrow}c^\dagger_{\bR,2\downarrow}
+ c^\dagger_{\bR,2\uparrow}c^\dagger_{\bR,1\downarrow}\big)\;|0\rangle,
\]
which is just a collection of singlets. In other words, in spite of being non-magnetic, our variational wavefunction is 
able to reproduce the correct super-exchange between dimers. This is a remarkable property that actually derives from 
the square-root dependence upon $d$ of $E_{hop}$, see Eq.~(\ref{Ehop-large-U}). If we considered a 
more conventional Gutzwiller operator $\mathcal{P}$ commuting with the single-particle density matrix, 
that amounts to further impose $2q = \sqrt{\delta_*^2 + \Delta_*^2}$, we would find $E_{hop}\propto d$, 
implying a transition into an unrealistic insulator with $d=0$ above a critical $U$.      
The obvious defect of the wavefunction is that, since it  
emphasizes strongly the role of individual dimers, the hopping among dimers, although finite for any $U$, is 
under-estimated with respect to the intra-dimer one. Therefore we do not expect the wavefunction to be particularly accurate 
for small $t_\perp/D$.

\section{Conclusions}
\label{Sec.IV}

In this work we have proposed an extension of the Gutzwiller variational approach to account for 
correlated models which display metal-insulator transitions into Mott insulators that escape any 
simple single-particle descriptions, like the Hartree-Fock approximation. The wavefunction has still the 
same form as the conventional Gutzwiller wavefunction,
\[
|\Psi_G\rangle = \mathcal{P}\, |\phi\rangle = \prod_\bR\, \mathcal{P}_\bR\, |\phi\rangle,
\]
with $\bR$ identifying unit cells that may also be non-primitive ones, with the novel feature that the operator 
$\mathcal{P}_\bR$ is non-hermitean and does not commute with the local single-particle density matrix. 
In essence, this property realizes a variational implementation  
of a Schrieffer-Wolff transformation~\cite{Schrieffer-Wolff}, although only restricted to the lattice sites 
within each unit cell.  
We have shown that, by slightly reducing the variational freedom, this wavefunction, like 
the conventional Gutzwiller wavefunction,~\cite{Gebhard} allows for an extension of the 
Gutzwiller approximation to evaluate average values, approximation that becomes 
exact in the limit of infinite coordination lattices.  

\begin{figure}
\includegraphics[width=8cm]{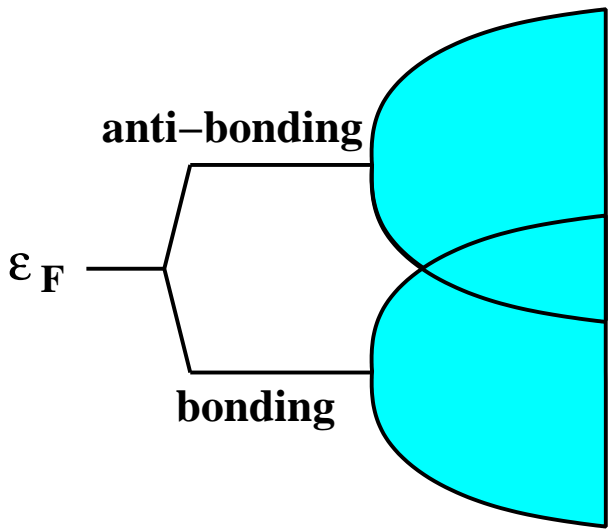}
\caption{The non-interacting density of states of the lattice of dimers. The bonding and anti-bonding state of each 
dimer give rise to two bands that overlap, leading to a metallic phase in the absence of interaction.
\label{dimer}}
\end{figure}

As an application, we have considered the Mott transition into a Peierls, or valence-bond, insulator, 
namely an insulator that is adiabatically connected to a collection of independent dimers. 
Such an insulator can not be 
described by Hartree-Fock simply because the singlet configuration of each dimer is not a Slater determinant. 
Specifically, we have considered the hypothetical situation shown in Fig.~\ref{dimer},
where the splitting between the bonding and antibonding orbitals of each dimer is assumed not to be sufficient 
to lead, in the absence of interaction, to a band insulator.  When interaction is taken into account, in the form of an on-site 
repulsion $U$, one expects, above a critical $U$, a transition from the metal into a Mott insulator. If 
magnetism is prevented, for instance by a sufficiently large splitting between bonding and anti-bonding orbitals and/or by frustration, 
the Mott insulator is non-magnetic.    
We have shown that our wavefunction overcomes the difficulties of Hartree-Fock theory and allows to study, albeit variationally, 
this transition. In particular we find that:  
\begin{itemize}
\item[({\sl i})] at the variational level the Mott transition is first order;
\item[({\sl ii})] the variational spectrum inside the Mott insulator looks similar to that of a Peierls insulator with a large 
hybridization gap, namely a large splitting between bonding and anti-bonding bands. In reality the gap is the Mott-Hubbard 
gap and the actual difference between the occupations of the bonding and 
anti-bonding bands, is small; 
\item[({\sl iii})] inter-site singlet-superconductivity appears around the transition.
\end{itemize}
While ({\sl i}) and ({\sl ii}) are presumably true, as they have been also found by more rigorous  
calculations~\cite{Moeller,Monien,biermann:dimer},
the emergence of superconductivity might be an artifact of the variational wavefunction.~\cite{Schiro'} 
Nevertheless, the possible occurrence of superconductivity is quite suggestive. 
It is known for instance that two-leg Hubbard ladders with nearest neighbor hopping display dominant superconducting fluctuations 
with the same symmetry that we find variationally~\cite{2chain}, although at half-filling they always describe 
non-magnetic spin-gaped insulators~\cite{Strong&Millis,shura} because of nesting. Moreover, the uncorrelated 
wavefunction $|\phi\rangle$ is quite similar to the wavefunctions 
used in Refs.~\onlinecite{Scalapino-RVB,sandro&dagotto} to simulate $t$-$J$ ladders. It would be surprising 
and interesting if this tendency towards superconductivity turned into a true symmetry breaking instability 
in higher dimensionality, as suggested by our analysis, which we think it is worth deserving further investigations. 

\noindent
{\sl Note added:} During the completion of this work, we became aware of a recent extension of slave-boson technique 
whose saddle-point solution closely resembles our variational approach.~\cite{Antoine} Indeed the two conditions 
we impose on the Gutzwiller operator, Eqs.~(\ref{cond-1}) and (\ref{cond-2}), are in one-to-one correspondence 
with the constraints identified in Ref.~\onlinecite{Antoine} within the slave-boson formalism.    

\begin{acknowledgments}
We are grateful to C. Castellani and E. Tosatti for their helpful comments and suggestions. We also thanks A. Georges 
for useful discussions in connection with Ref.~\onlinecite{Antoine}.
\end{acknowledgments}

\appendix*
\section{The natural basis} 
\label{The natural basis}

Let us assume that, in the Nambu-spinor representation 
\[
\left(
\begin{array}{c}
c^\dagga_{e\uparrow}\\
c^\dagger_{e\downarrow}\\
\end{array}
\right),\;
\left(
\begin{array}{c}
c^\dagga_{o\uparrow}\\
c^\dagger_{o\downarrow}\\
\end{array}
\right),
\]
the uncorrelated wavefunction has the following density matrices  
\be
\hat{C}_e = 
\left(
\begin{array}{cc}
n_e/2 & \Delta_{SC}/2 \\
\Delta_{SC}/2 & 1 - n_e/2 \\
\end{array}
\right),\;
\hat{C}_o = 
\left(
\begin{array}{cc}
n_o/2 & -\Delta_{SC}/2 \\
-\Delta_{SC}/2 & 1 - n_o/2 \\
\end{array}
\right).\label{APP:C}
\ee
The natural orbitals are obtained by the unitary transformation
\ba
d^\dagga_{e(o)\uparrow} &=& \cos\theta_{e(o)} \,c^\dagga_{e(o)\uparrow} + \sin\theta_{e(o)} \,c^\dagger_{e(o)\downarrow}\\
d^\dagga_{e(o)\downarrow} &=& \cos\theta_{e(o)} \,c^\dagga_{e(o)\downarrow} - \sin\theta_{e(o)} \,c^\dagger_{e(o)\uparrow},
\ea
where 
\be
\tan 2\theta_{e} = \frac{\Delta_{SC}}{n_e-1},\qquad \tan 2\theta_o = \frac{-\Delta_{SC}}{n_o-1},
\label{APP:theta}
\ee
and posses a diagonal density matrix with the non-vanishing elements given by 
\be
\langle \phi|\, d^\dagger_{e(o)\sigma} d^\dagga_{e(o)\sigma}\,|\phi\rangle = 
\frac{1}{2} + q_{e(o)},
\label{APP:C-natural}
\ee
where 
\be
q_{e(o)} = \frac{1}{2}\,\sqrt{(n_{e(o)}-1)^2 + \Delta_{SC}^2}.
\label{APP:q}
\ee
In the natural basis we introduce states that have the same formal expression as in the original basis but are built with 
$d$-operators, and denote them as $|\wt{n},\Gamma\rangle$. The transformation rules from these states 
to the original ones are
\bea
|0\rangle &=& \cos\theta_e\,\cos\theta_o |\widetilde{0}\rangle + 
\sin\theta_e\,\sin\theta_o\,|\widetilde{4}\rangle \nonumber \\
&& + \cos\theta_e\,\sin\theta_o\,|\widetilde{2},o\rangle 
+ \sin\theta_e\,\cos\theta_o\,|\widetilde{2},e\rangle,  \nonumber \\
|1,e(o),\sigma\rangle &=& \cos\theta_{o(e)}\,|\widetilde{1},e(o),\sigma\rangle + 
\sin\theta_{o(e)}\, |\widetilde{3},e(o),\sigma\rangle, \nonumber\\
|2,e(o)\rangle &=& \cos\theta_e\,\cos\theta_o\, |\widetilde{2},e(o)\rangle + 
\cos\theta_{e(o)}\,\sin\theta_{o(e)}\,|\widetilde{4}\rangle \nonumber\\
&& -\sin\theta_{e(o)}\,\cos\theta_{o(e)}\,|\widetilde{0}\rangle \nonumber \\
&& - \sin\theta_{e(o)}\,\sin\theta_{o(e)}\, |\widetilde{2},o(e)\rangle, \nonumber\\
|2,1,S_z\rangle &=& |\widetilde{2},1,S_z\rangle,\label{APP:transform}\\
|2,0\rangle &=& |\widetilde{2},0\rangle, \nonumber\\
|3,e(o),\sigma\rangle &=& \cos\theta_{o(e)}\,|\widetilde{3},e(o)\rangle 
- \sin\theta_{o(e)}\,|\widetilde{1},e(o)\,\sigma\rangle,\nonumber\\
|4\rangle &=& \cos\theta_e\,\cos\theta_o |\widetilde{4}\rangle + 
\sin\theta_e\,\sin\theta_o\,|\widetilde{0}\rangle \nonumber \\
&& - \cos\theta_e\,\sin\theta_o\,|\widetilde{2},e\rangle 
- \sin\theta_e\,\cos\theta_o\,|\widetilde{2},o\rangle .\nonumber
\eea
The inverse transformation is obtained by letting $\theta_{e(o)}\to -\theta_{e(o)}$. 

The hopping operator in the original representation can be written as 
\bea
\mathcal{H}_\perp &=& -2t_\perp\,\sum_\sigma\, c^\dagger_{1\sigma}c^\dagga_{2\sigma} + H.c. \nonumber \\
&=& 
-2t_\perp\,\sum_\sigma\, c^\dagger_{e\sigma}c^\dagga_{e\sigma} - c^\dagger_{e\sigma}c^\dagga_{e\sigma} \nonumber \\
&=& -2t_\perp\,\Bigg[\sum_\sigma\, \Pj{3,o,\sigma}{3,o,\sigma} - \Pj{3,e,\sigma}{3,e,\sigma} \nonumber \\
&& + \Pj{1,e,\sigma}{1,e,\sigma} - \Pj{1,o,\sigma}{1,o,\sigma} \nonumber \\
&& + 2\,\Pj{2,e}{2,e} - 2\,\Pj{2,o}{2,o}\Bigg], \label{APP:hopping}
\eea
while the interaction operator as 
\bea
\mathcal{H}_U &=&\frac{U}{2}\sum_{i=1}^2\, \left(n_i-1\right)^2 = \frac{U}{2}\,\Bigg[2\,\Big(\Pj{0}{0} + \Pj{4}{4}\Big)\nonumber\\
&& + \sum_\sigma\, \sum_{n=1,3}\, \Pj{n,e,\sigma}{n,e,\sigma} + \Pj{n,o,\sigma}{n,o,\sigma}\nonumber\\
&& + \Big(|2,e\rangle + |2,o\rangle \Big)\,\Big(\langle 2,e|+ \langle 2,o|\Big) \nonumber\\
&& + 2\,\Pj{2,0}{2,0}\Bigg].\label{APP:U}
\eea 
Their expression in the natural basis can be obtained by the transformation rules (\ref{APP:transform}).


\end{document}